\documentclass[prb,twocolumn,amsmath,letterpaper]{revtex4}
\usepackage[dvips]{color}
\usepackage{graphicx}
\usepackage{verbatim}

\begin{document}
\title{Single-Molecule Device Prototypes for Protein-Based Nanoelectronics:
Negative Differential Resistance and Current Rectification in Oligopeptides}
\author{David M. Cardamone and George Kirczenow}
\affiliation{Physics Department, Simon Fraser University,
Burnaby, BC V5A 1S6, Canada}

\begin{abstract}
We investigate electrical conduction through individual oligopeptide molecules
thiol-bonded between gold nanocontacts
using \emph{ab initio} and semi-empirical techniques. Our theory
explains for the first time these molecules' experimentally observed
current-voltage characteristics, including
both the magnitude and
rectification of the current, and uses no adjustable parameters. We identify the mechanism of the observed
current rectification, and predict that it will result in 
negative differential
resistance at moderate biases. Our findings open the way to the realization
of protein-based nanoelectronic devices.
\end{abstract}

\maketitle

\section{Introduction}

Living cells survive and function through their unique ability to
manufacture a vast catalog of specific molecules repeatably and reliably,
a capacity long recognized to hold promise for the design of
nanoelectronic devices.\cite{gilmanshin93}
Whereas much experimental and theoretical work has
been devoted to understanding the electron transport properties of single
nucleic acid molecules, relatively little has focused on the same qualities of
other biomolecules. In particular, over the course of its life a
cell produces far more polypeptide chains than DNA sequences, while
amino acids, the building blocks of proteins, display five times more 
diversity
than DNA bases. As such, the single-molecule electron transport
characteristics of these protein fragments, and of entire proteins, is
a field with great potential, which
is only beginning to be explored. 

Recently,
experimentalists have succeeded in
addressing the effects of primary protein structure,
\emph{i.e.}\ 
specific amino acid sequences, on conductance.\cite{xiaojacs04,xiaoac04} These experiments, applying the 
STM-break junction (STMBJ) technique \cite{xuscience03,xujacs03,xiaonl04}
to oligopeptide molecules,
present an exciting new approach to
biologically-based nanoelectronic devices:
They offer 
a broad class of
well defined experimental systems
operating at
length scales on which coherent quantum effects are dominant. Reference \onlinecite{xiaoac04}, in particular, demonstrated this promise with
non-equilibrium results evincing a striking current rectification.

\begin{figure}
\setlength{\unitlength}{\columnwidth}
\begin{picture}(1,1)
\put(.915,.257){\color{blue}\line(-4,1){.3}}
\put(.35,.37){\includegraphics[keepaspectratio=true,width=.15\unitlength,angle=270]{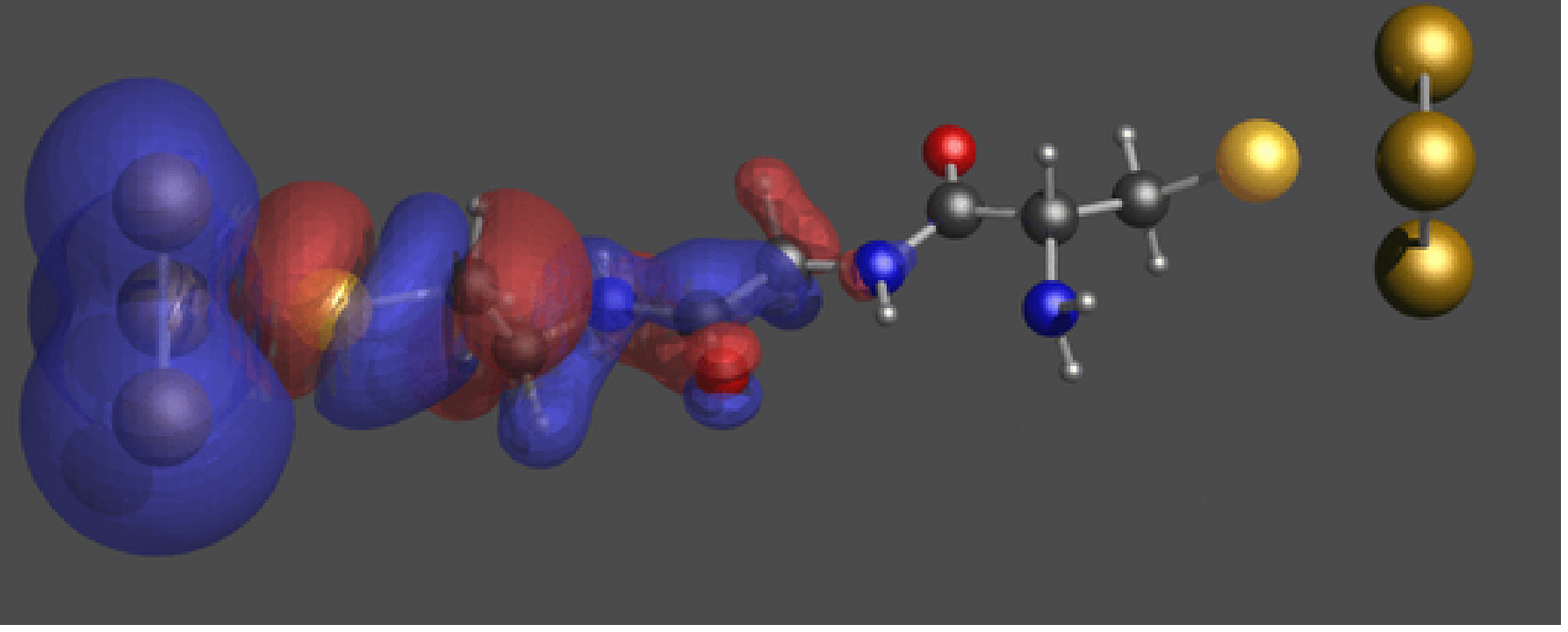}}
\put(0,0){\includegraphics[keepaspectratio=true,width=\unitlength]{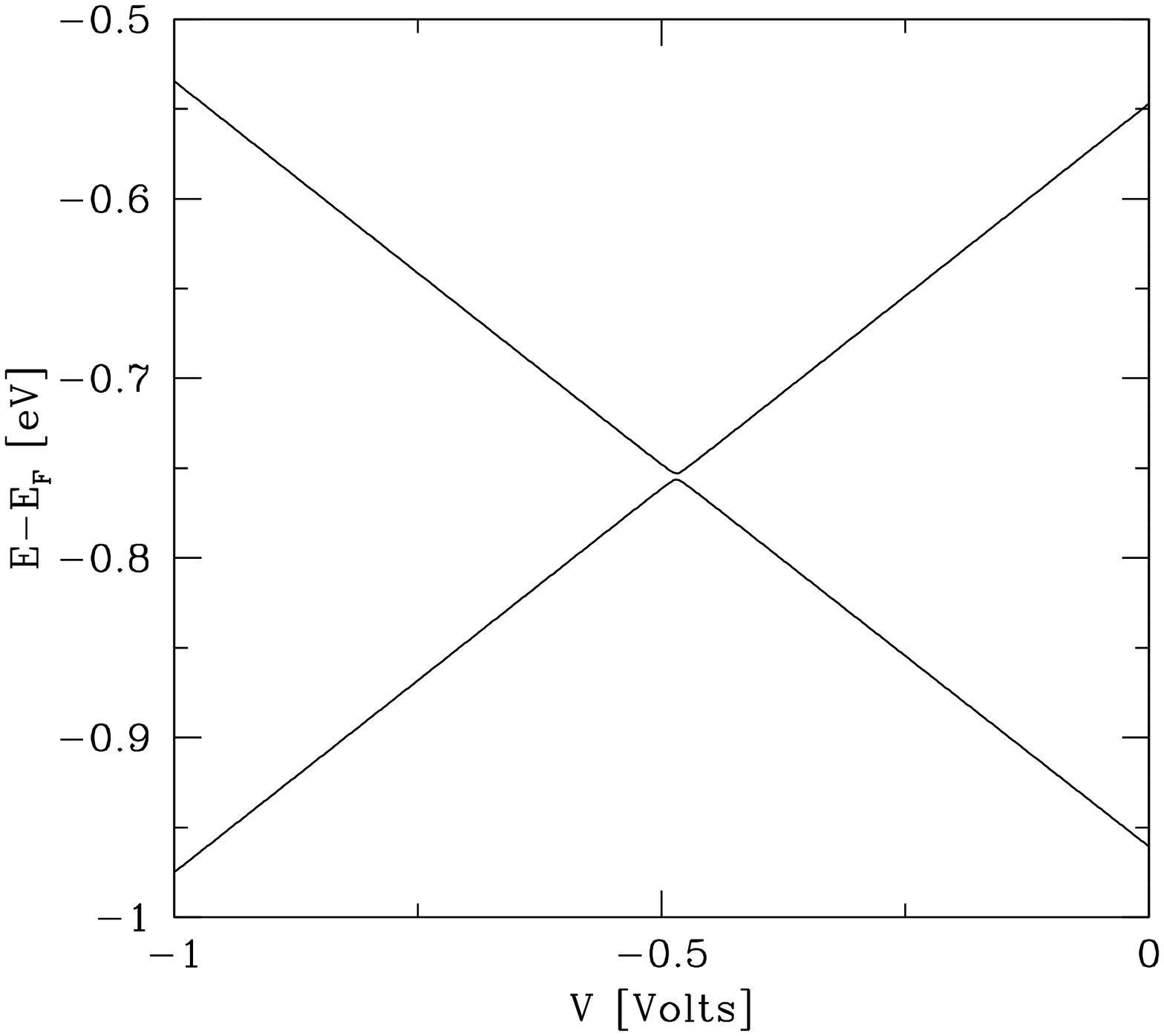}}
\put(.915,.257){\circle*{.01}}
\put(.915,.85){\color{blue}\line(-4,-1){.3}}
\put(.35,.87){\includegraphics[keepaspectratio=true,width=.15\unitlength,angle=270]{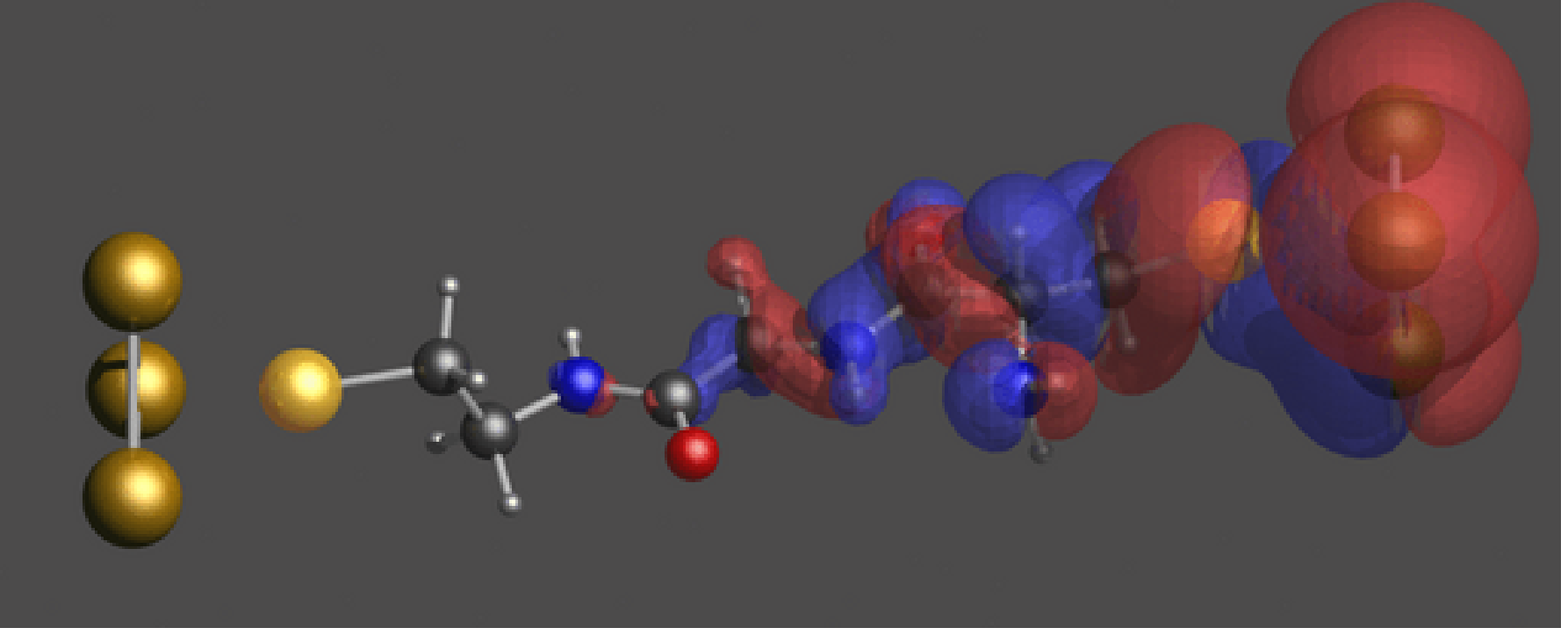}}
\put(.915,.85){\circle*{.01}}
\put(.541,.555){\color{blue}\line(-1,0){.3}}
\put(.541,.555){\circle*{.01}}
\put(.541,.545){\color{blue}\line(1,0){.3}}
\put(.541,.545){\circle*{.01}}
\put(.17,.6){\includegraphics[keepaspectratio=true,width=.1\unitlength,angle=270]{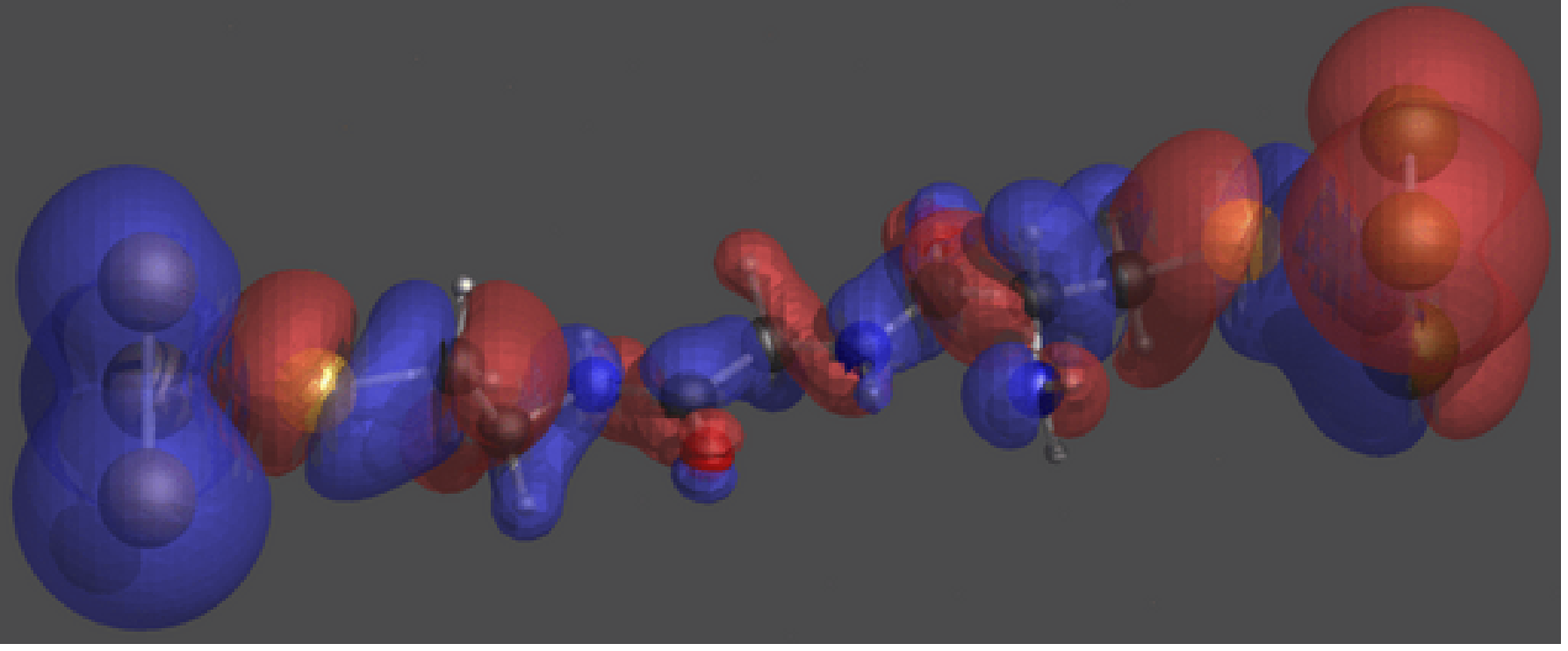}}
\put(.65,.6){\includegraphics[keepaspectratio=true,width=.1\unitlength,angle=270]{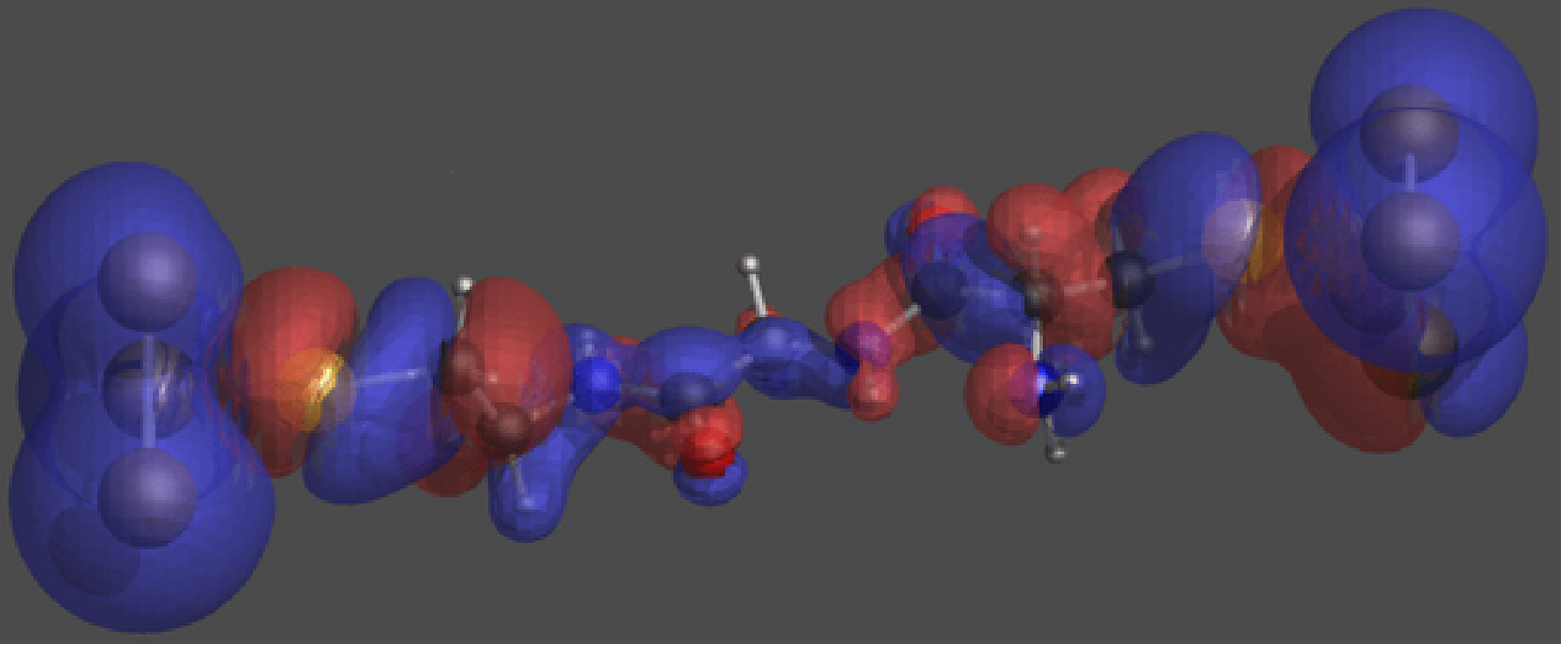}}
\put(.8,.73){$R$}
\put(.8,.36){$L$}
\end{picture}
\caption{(Color online) Anticrossing of generalized eigenstates $R$ and $L$ of the
  simplified extended molecule discussed in the text for cysteamine-Gly-Cys
  bonded to two hollow sites in the Au(111) leads. In the full system, the
  anticrossing's abruptness is tempered by broadening from the leads. The
  insets show the states at specific bias voltages.\cite{amountstretch}}
\label{anticrossing}
\end{figure}

The purpose of this article is to present the first theoretical treatment of
electron transport through these systems. We explain the
phenomena observed in these experiments, in particular the 
current rectification, by use of
Landauer conductance theory,\cite{datta} and demonstrate 
good quantitative agreement with the experimental
data. 
We show that the
rectification is the result of resonance between gold-molecule
interfacial states partially localized on either side of the molecule (see Fig.~\ref{anticrossing})\nocite{amountstretch}; these states are
detuned from each other in equilibrium due to the oligopeptides' intrinsic
asymmetry. 
Moreover, we predict that, at moderately higher voltages, a related mechanism will generate
negative differential resistance (NDR),
\emph{i.e.,} a decrease in current with increased bias voltage;
this phenomenon has important device applications, 
including low-noise amplification,\cite{vanderzielsse83}
high frequency oscillators,\cite{brownapl91} analog-to-digital
conversion,\cite{broekaertieeejssc98} and digital
logic.\cite{mathewspieee99} Thus, our findings open the way to the
realization of practical oligopeptide, and ultimately protein-based,
nanoelectronic devices.

We organize the paper as follows. Section \ref{sec:model} presents the
theoretical formalism we use to model charge transport through
metal--oligopeptide--metal junctions. In Section \ref{sec:results}, we discuss
our results for these systems, including determination of the experimental
geometry, explanation of their rectifying properties, and prediction of NDR. Section
\ref{sec:conclusion} summarizes our main conclusions.

\section{\label{sec:model}Model}

To determine the molecular geometry in the presence of the gold leads, we apply a
series of density functional theory \emph{ab
  initio} calculations to relax the molecule between two fixed gold
atoms.\cite{gaussian} At each step, the distance between the two atoms is
increased slightly, and the geometry relaxed again, to simulate the stretching
the molecule undergoes in an STMBJ experiment. For calculations reported here, the
molecules have been stretched to their greatest length before breakage occurs,
although we find that less stretched molecules have essentially the same
conductance features and order of magnitude. In the case of hollow binding
sites, we replace the gold atom with the three nearest-neighbor gold atoms,
and relax the molecule one final time.

To determine conductance and current, we first partition the \mbox{metal--oligopeptide--metal} system into left and right
macroscopic gold electrodes, and an extended molecule containing
the central organic molecule 
as well as 50-100 atoms of each Au(111)
contact.\cite{broadeningnote} The Hamiltonian of the full system is therefore the sum of 
two terms:
\begin{equation}
H=H_{mol}+H_{leads}.
\end{equation}
$H_{mol}$ 
describes
the isolated extended molecule,
and $H_{leads}$ does the same for the macroscopic electrodes, as well as containing
the quantum mechanical coupling between them and the extended molecule. 

We employ a
semi-empirical model for $H_{mol}$; such models are known to be highly
effective in predicting, reproducing, and explaining experimental results for
metal--organic molecule--metal junctions (\emph{e.g.}, Ref.~\onlinecite{dattaprl97},
Ref.~\onlinecite{emberlyprb01} and the results of Refs.~\onlinecite{reedscience97} and \onlinecite{xiaonl04}, Ref.~\onlinecite{kushmerickprl02}). 
The Hamiltonian of the full
extended molecule is given by extended H\"uckel
theory.\cite{hoffmanjcp63,ammeterjacs78,cerdaprb00,kirczenowprb05,kienlejap06,bind}

$H_{leads}$ provides both decoherent broadening and elastic
scattering to the discrete states of the extended molecule. To model this,
we assign
semi-infinite, one-dimensional
leads~\cite{mujicajcp94II} to 
each gold orbital not immediately adjacent to the oligopeptide; their number is
equal to the coordination number of gold minus the number of nearest-neighbor
atoms included in the extended molecule. We take the intersite hopping within
each lead, as well as the hopping between lead and orbital, to be 5eV.
These choices 
are made to 
quantitatively reproduce the
quantized conductance~\cite{hansenapl00,mehrezprb02} of a gold quantum point
contact, and are not fitted to any molecular electronics experiment.

The open system described by $H$ requires an infinite Fock space.
To solve 
the full molecule-plus-leads system, then, we make use of Dyson's Equation,
which gives the retarded Green function 
\begin{equation}
\label{dyson}
G(E)=\left[G_{mol}^{-1}(E)-\Sigma(E)\right]^{-1}.
\end{equation}
Here
\begin{equation}
G_{mol}(E)=\frac{1}{SE-H_{mol}+i0^+}
\end{equation}
is the Green function for the extended molecule only,\cite{emberlyprl98}
and the 
retarded self-energy 
$\Sigma(E)$ is given by 
Ref.~\onlinecite{mujicajcp94II}.
$S$ is the atomic orbital overlap matrix of the extended molecule.
Equation~(\ref{dyson}) thus represents
an exact method to incorporate the
effects of $H_{leads}$ into transport.

\begin{table*}
\caption{Comparison of the conductances found in our model\cite{amountstretch} (in units of $g_0$)
  for various bonding geometries
  with those of
  the STMBJ experiments.\cite{xiaojacs04,xiaoac04} In the column
  headings, the first entry 
  indicates the bonding geometry on the amine side of the molecule, the second the other. Only hollow bonding at both interfaces is compatible with experiment.}
\label{tab}
\begin{tabular}{|l|r@{.}l|r@{.}l|r@{.}l|r@{.}l|r@{.}l|}
\hline
molecule & \multicolumn{2}{c}{experimental}\vline &
\multicolumn{2}{c}{hollow-hollow}\vline & \multicolumn{2}{c}{hollow-top}\vline
& \multicolumn{2}{c}{top-hollow}\vline & \multicolumn{2}{c}{top-top}\vline \\
\hline\hline
cysteamine-Cys & 1&8$\times 10^{-4}$ & 5&4$\times 10^{-5}$& 1&5$\times
10^{-6}$ & 9&6$\times 10^{-7}$ & 5&8$\times 10^{-9}$\\ \hline
cysteamine-Gly-Cys & 4&2$\times 10^{-6}$ & 4&3$\times 10^{-6}$ & 5&3$\times
10^{-8}$ & 3&7$\times 10^{-8}$ & 1&5$\times 10^{-9}$ \\ \hline
Cys-Gly-Cys & 5&3$\times 10^{-6}$ & 4&2$\times 10^{-6}$ & 9&3$\times 10^{-8}$&
1&0$\times 10^{-7}$ & 1&7$\times 10^{-9}$\\ \hline
cysteamine-Gly-Gly-Cys & 5&0$\times 10^{-7}$ & 4&1$\times 10^{-8}$& 7&3$\times
10^{-10}$& 1&9$\times 10^{-10}$& 2&2$\times 10^{-11}$\\ \hline
\end{tabular}
\end{table*}

We model electrostatic effects, including the potential imposed by the
contacts as well as 
screening by the free ionic charges in the electrolyte solution surrounding
the molecule, with a linearized Debye--H\"uckel
theory.\cite{landau} 
In such a theory, 
Poisson's equation for the total electric potential
$\Phi({\bf r})$ becomes
\begin{equation}
\label{DH}
\nabla^2\Phi({\bf r})=\frac{2n\mathrm{e}^2}{k_B\mathrm{T}\epsilon}\Phi({\bf r}),
\end{equation}
where $n=.1\mathrm{M}$ is the concentration of the solution, $-\mathrm{e}$ is
the charge of an electron, $\epsilon$ is
the dielectric constant of water, $k_B$ is the Boltzmann constant, and $\mathrm{T}$ is room
temperature.\cite{xiaojacs04suppl}

We solve Eq.~(\ref{DH}) for parallel-plate boundary conditions, where the
plates are taken to lie at the metal-molecule interfaces, and take $\Phi({\bf
  r})$ constant within the metal nanoclusters themselves.
In the electrostatic gauge for which the voltages of the left and
right lead add to zero, the solution of Eq.~(\ref{DH}) in the region
containing the molecule is
\begin{equation}
\label{Phi}
\Phi(z)=\frac{V}{2}\left[\cosh\left(\frac{z}{\lambda}\right)-\frac{\cosh\left(\frac{d}{\lambda}\right)+1}{\sinh\left(\frac{d}{\lambda}\right)}\sinh\left(\frac{z}{\lambda}\right)\right],
\end{equation}
where $z=0$ is one boundary surface, $z=d$ the other, $V$ the applied bias, and the Debye screening
length is $\lambda=\sqrt{k_B\mathrm{T}\epsilon/2n\mathrm{e}^2}$. For the
experimental systems of
Refs.~\onlinecite{xiaojacs04} and \onlinecite{xiaoac04},
$\lambda=0.97\mathrm{nm}$; since this length is greater than $d/2$
for all the metal--molecule--metal junctions studied here,
the effect of screening in these systems is slight, and our results are
robust against reasonable changes in the experimentally determined parameters $\mathrm{T}$, $\epsilon$, $d$,
and $n$. 
It should also be noted that,
although
Eq.~(\ref{Phi}) is written as an example in a particular gauge, the results of
our calculations are gauge invariant.

We use a simple ansatz to include the electrostatic effect of $\Phi$ into
$H_{mol}$: \cite{emberlycp02}
\begin{equation}
\left(H_{mol}\right)_{ij}\rightarrow
\left(H_{mol}\right)_{ij}-\mathrm{e}S_{ij}\frac{\Phi\left({\bf
      r}_i\right)+\Phi\left({\bf r}_j\right)}{2},
\end{equation}
where, ${\bf r}_i$ is the position of the $i^{th}$ atomic orbital. Thus, each 
diagonal element is gated in the usual way, and off-diagonal elements are
modified according to the non-orthogonality of the basis states, as is typical
in a H\"uckel theory. In addition, of course, 
the site energies of each lead and metallic nanocluster shift
uniformly with their respective metallic contact's electrostatic
potential. All the quantities in Eq.~(\ref{dyson}) thus become functions of the
applied bias $V$, without the use of free parameters.

The Landauer formalism \cite{datta} yields the conductance and current through the
molecule. The probability for an electron of energy $E$ to move from the left
set of leads to the right is
\begin{equation}
T(E,V)=\!\!\!\!\sum_{\substack{\alpha\in\mathrm{left}\\ \beta\in\mathrm{right}}}\!\!\!\!\mathrm{Tr}\left[\Gamma^{(\beta)}(E,V)G(E,V)\Gamma^{(\alpha)}(E,V)G^\dagger(E,V)\right],
\end{equation}
where
\begin{equation}
\Gamma^{(\alpha)}(E,V)\equiv -2\,\mathrm{Im}\left[\Sigma^{(\alpha)}(E,V)\right]
\end{equation}
is the broadening function for the term of $\Sigma$ from lead $\alpha$.
The zero-bias conductance of the metal--molecule--metal junction is $g=g_0T(E_F,0)$,
with the conductance quantum $g_0=2\mathrm{e}^2/h$ and the bulk gold Fermi energy $E_F$
evaluated 
for a 459-atom spherical gold nanocluster. The current is~\cite{datta}
\begin{equation}
I(V)=\frac{\mathrm{e}}{h}\int_{-\infty}^\infty
dE\, T(E,V)\left[f(E-\mu_R)-f(E-\mu_L)\right],
\end{equation}
where $f(E-\mu_{R,L})$ is the Fermi function of the leads on the right (left)
side of the extended molecule, and $\mu_{R,L}(V)$ is the corresponding electrochemical potential.

\section{\label{sec:results}Results and Discussion}

\subsection{Conductance and Bonding Geometry}

\begin{figure}
\setlength{\unitlength}{.3\columnwidth}
\begin{picture}(3,4)
\put(1.93,2.97){\includegraphics[keepaspectratio=true,width=.8\unitlength]{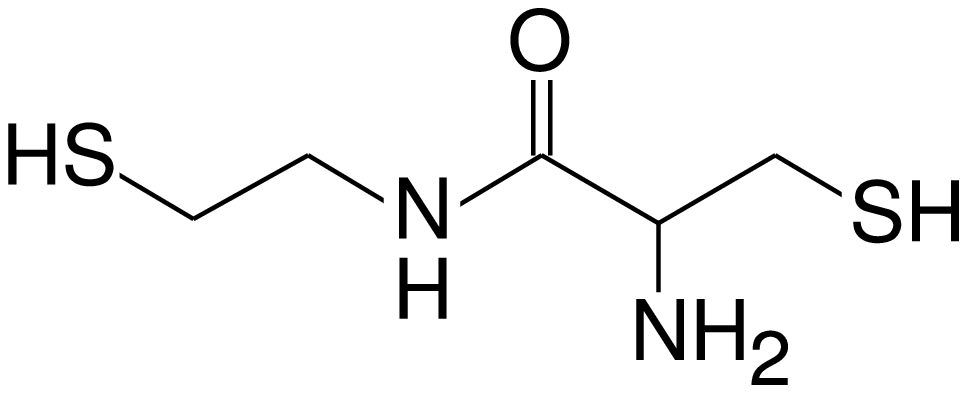}}
\put(1.28,2.54){\includegraphics[keepaspectratio=true,width=1.05\unitlength]{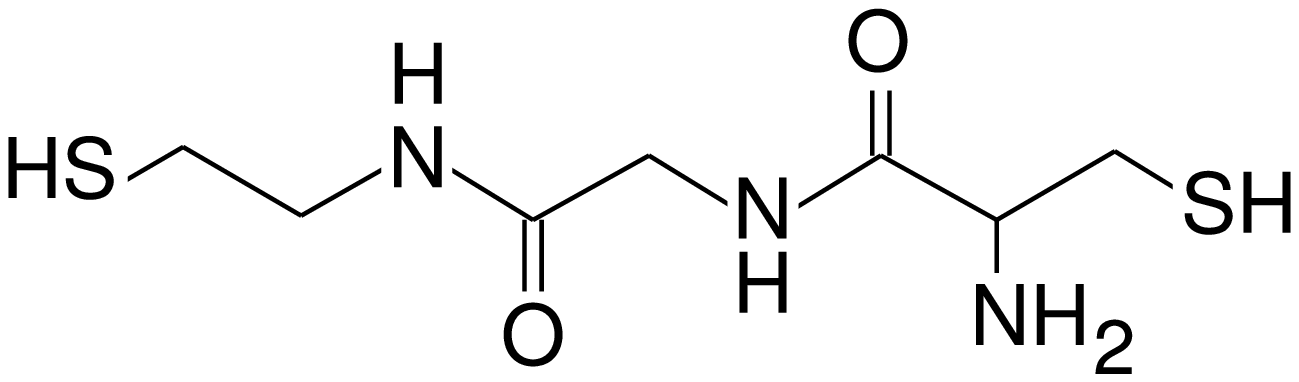}}
\put(1.1,1.48){\includegraphics[keepaspectratio=true,width=1.05\unitlength]{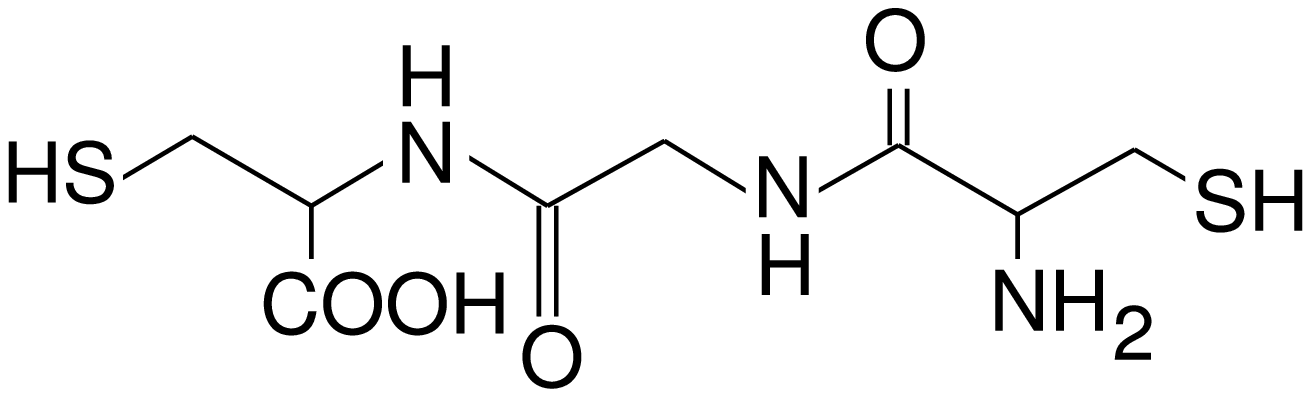}}
\put(1.05,.54){\includegraphics[keepaspectratio=true,width=1.25\unitlength]{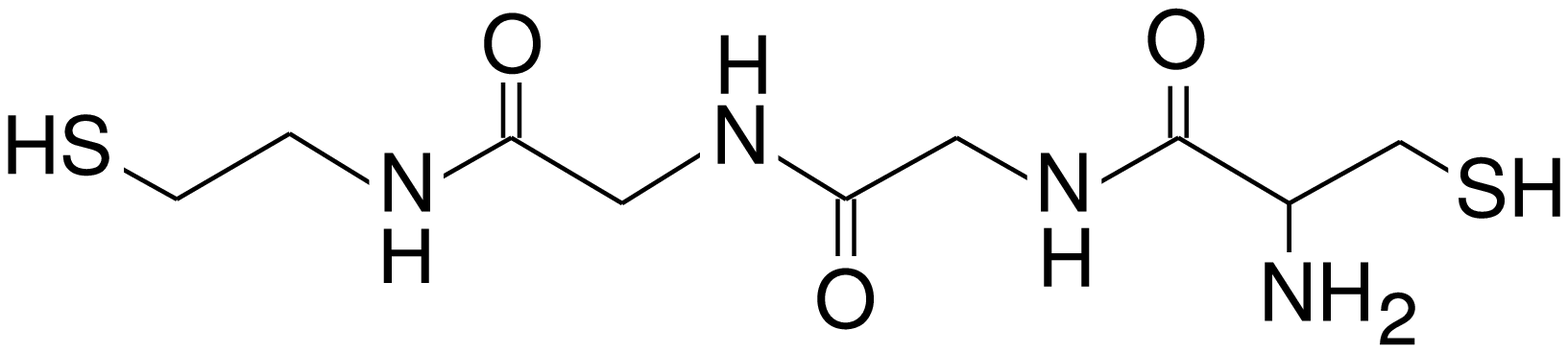}}
\put(0,0){\includegraphics[keepaspectratio=false,width=3\unitlength,height=\unitlength]{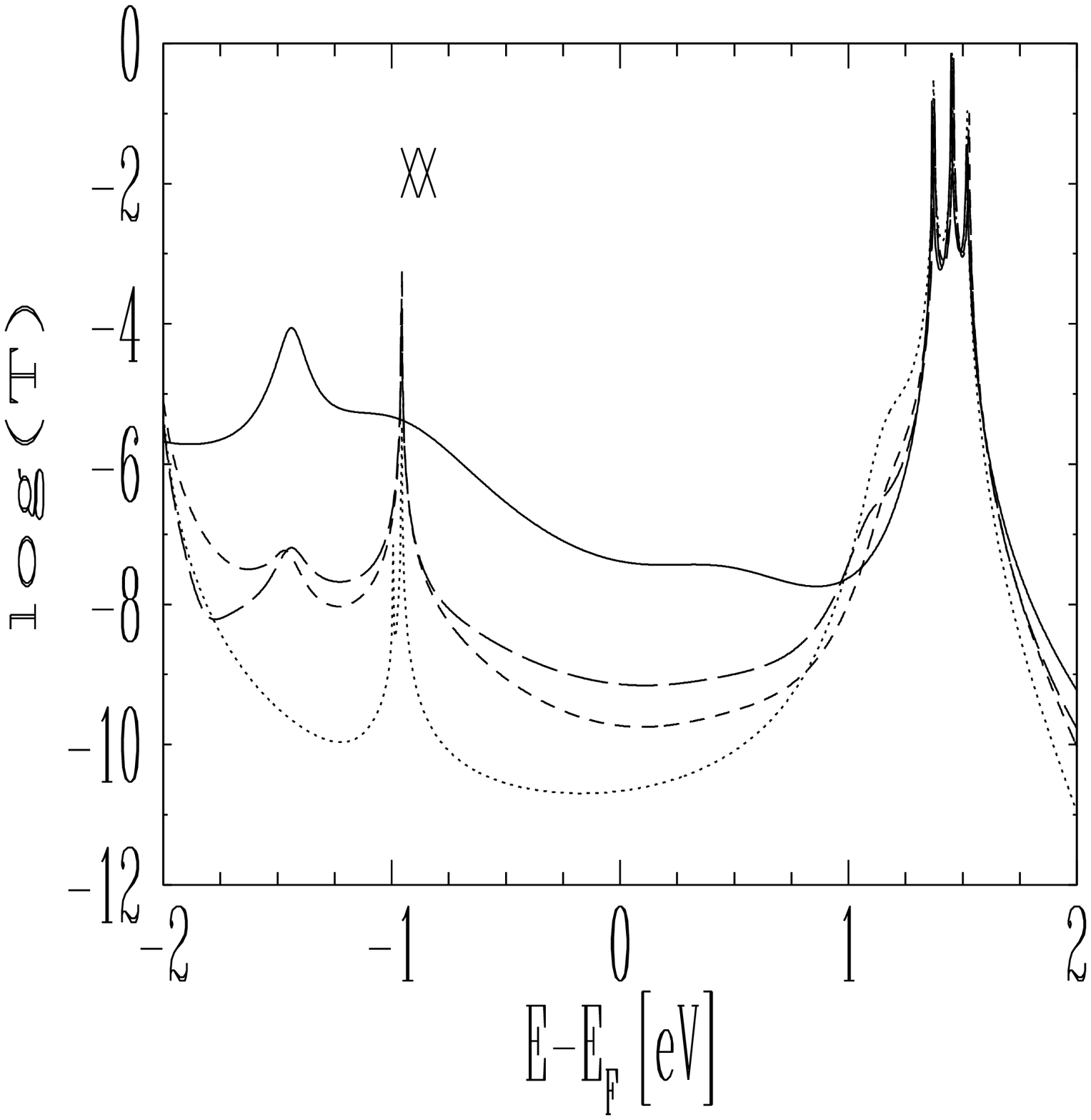}}
\put(0,1){\includegraphics[keepaspectratio=false,width=3\unitlength,height=\unitlength]{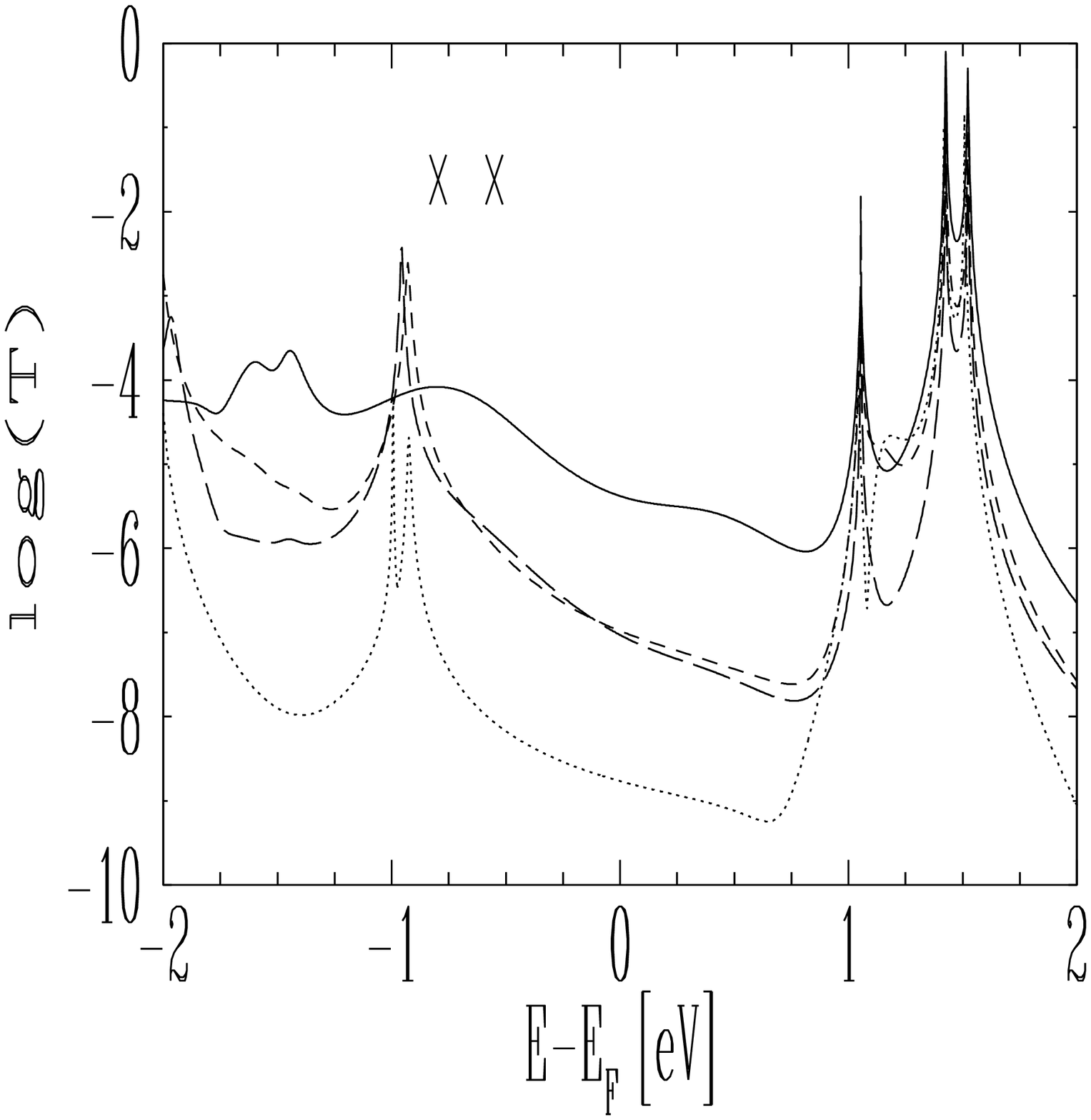}}
\put(0,2){\includegraphics[keepaspectratio=false,width=3\unitlength,height=\unitlength]{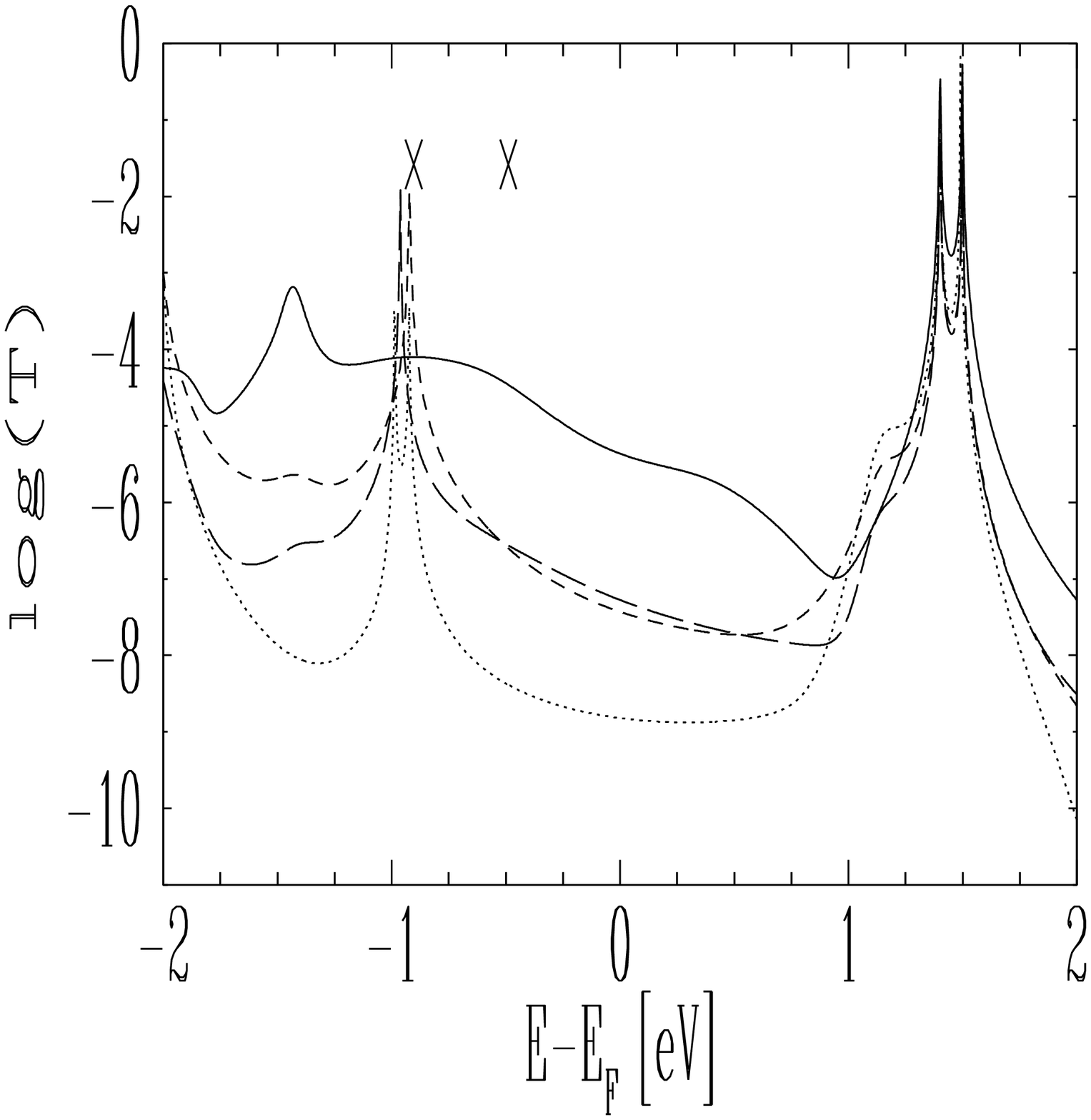}}
\put(0,3){\includegraphics[keepaspectratio=false,width=3\unitlength,height=\unitlength]{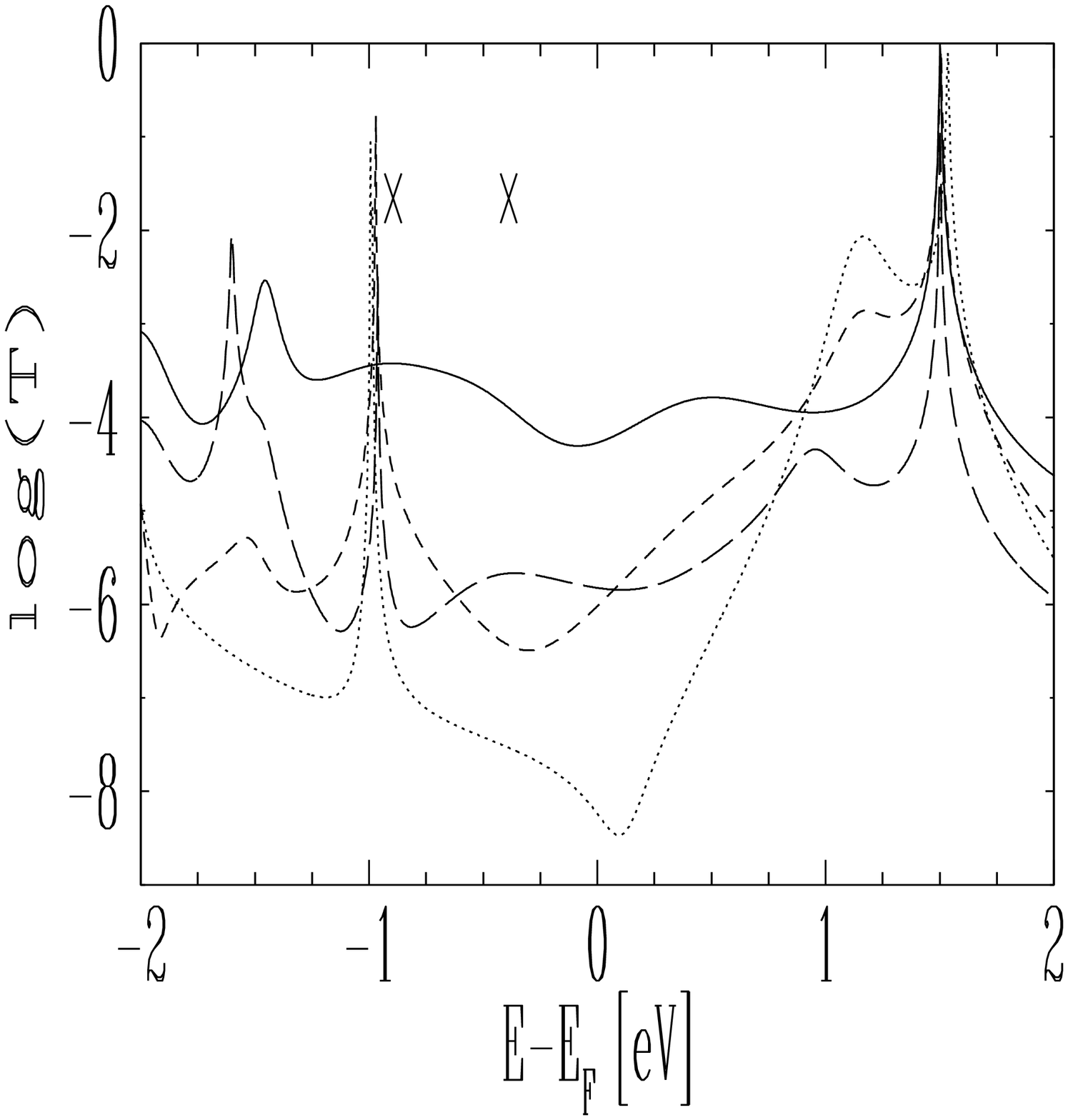}}
\put(.5,3.25){\scriptsize cysteamine-Cys}
\put(.5,2.25){\scriptsize cysteamine-Gly-Cys}
\put(.5,1.25){\scriptsize Cys-Gly-Cys}
\put(1.6,0.23){\scriptsize cysteamine-Gly-Gly-Cys}
\put(1.08,3.81){$L$}
\put(1.37,3.81){$R$}
\put(1.08,2.83){$L$}
\put(1.31,2.83){$R$}
\put(1.02,1.83){$L$}
\put(1.275,1.83){$R$}
\put(.95,.81){$L$}
\put(1.1,.81){$R$}
\end{picture}
\caption{Equilibrium transmission functions\cite{amountstretch} $T(E,0)$ for various bonding geometries for each
  of the four oligopeptide molecules studied experimentally in Refs.~\onlinecite{xiaojacs04} and \onlinecite{xiaoac04}. Solid lines indicate hollow-site
  bonding geometry on both sides. Dotted: top-site at both metal-molecule interfaces. Long dashes:
  hollow-site bonding for the sulfur on the amine-terminated side of the
  molecule and top-site bonding for the other. Short dashes: \emph{vice
    versa}. Experimental conductances agree well with the solid curves in all
  cases. For the hollow-hollow geometry, the energies of the two molecular
  orbitals 
  responsible for rectification
  are indicated with crosses. Inset: Structural
formulae of the gas-phase molecules. When bonded to gold, the terminating
hydrogen atoms are lost.}
\label{eq}
\end{figure}

We studied the conductance and current of
each oligopeptide molecule measured experimentally
in Refs.~\onlinecite{xiaojacs04} and \onlinecite{xiaoac04}, varying several
parameters of the experimental geometry in each case. These included 
the sulfur-to-sulfur distance as the molecule was stretched, 
the presence of
additional molecules of the self-assembled monolayers, and bond geometry;
only the last was found to have an
impact on the observables of more than a factor of order unity. 
Figure \ref{eq} demonstrates this
dependence, considering all four possibilities involving the two most likely
bonding geometries of the sulfur:\cite{lijacs06} either directly to a single gold atom on top of
the Au(111) substrate, or above a triangular hollow. Table \ref{tab}
compares the conductances obtained from our model\cite{amountstretch} with experiment. 

The experimental data are
consistent only with the model calculations for a hollow bonding geometry at
both ends of each oligopeptide molecule. Since this model is known to produce
quantitatively correct conductance in other gold-thiol systems,\cite{dattaprl97,emberlyprb01,kushmerickprl02,alkanenote} we conclude that this
is indeed the experimental geometry, and that the orders-of-magnitude smaller
conductances associated with 
top-bound geometries and 
with molecules
bound via monatomic gold chains\cite{kruegerprl02} form part of the low-conductance noise in the
experimental histograms.\cite{xiaojacs04} 
In this way, the 
peaks of the conductance 
histograms of the
STMBJ technique 
correspond to
a specific geometry for the metal--molecule--metal junction,
despite sulfur's ability to bond to a gold substrate in several geometries.

Further evidence that the single oligopeptide molecules of the experiments are
bound to hollow substrate sites at each end is found in the
behavior of conductance as a function of molecular stretching. Using DFT
relaxation at progressively larger lead-lead distances, we find that internal
degrees of freedom within
the oligopeptide molecule itself accommodate only about one third of the total
stretching induced by pulling the leads apart. The remaining stretching is
allowed only by repositioning of the entire molecule between the two metal
contacts, i.e. by changes of the Au-S-C bond angles and the dihedral angles
involving both gold contacts and the molecule.
This finding is consistent with the
persistence lengths of polypeptide chains, typically several nanometers or
more.

\begin{figure}
\setlength{\unitlength}{\columnwidth}
\begin{picture}(2,1.45)
  \put(0,.45){\includegraphics[keepaspectratio=true,width=\unitlength]{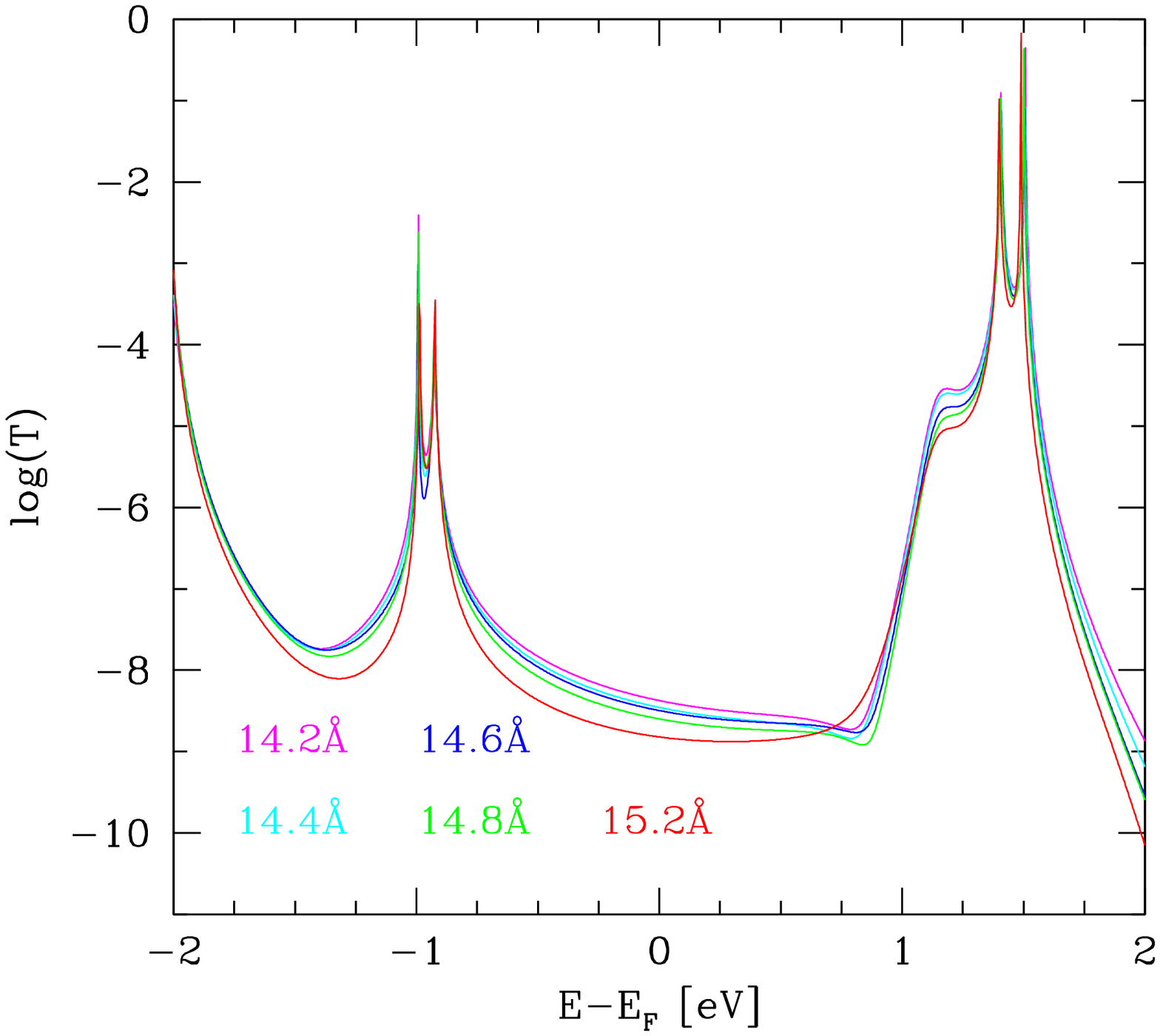}}
  \put(.37,.97){\includegraphics[keepaspectratio=true,width=.35\unitlength]{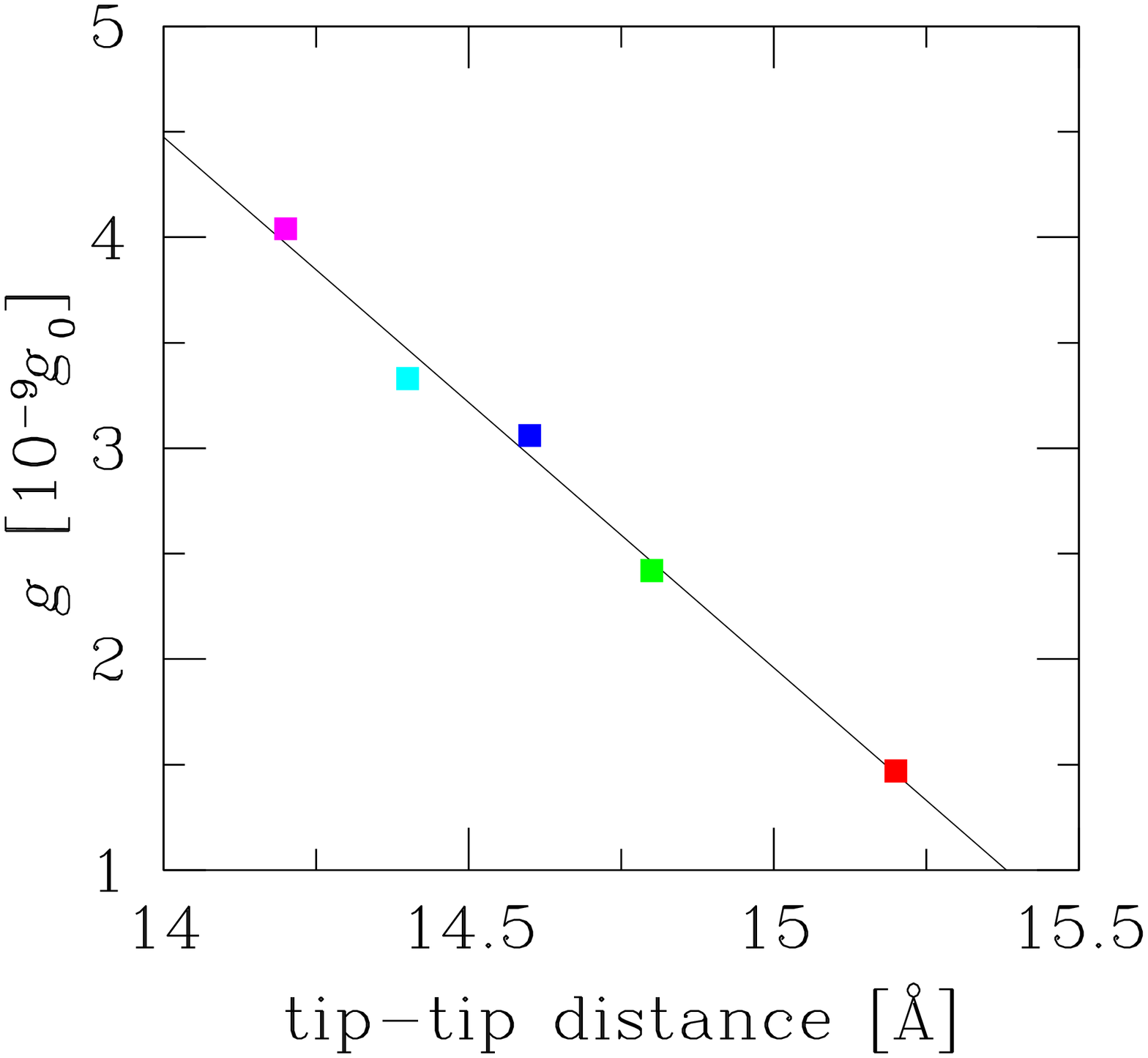}}
  \put(.15,1.32){(a)}
  \put(0,.26){\includegraphics[keepaspectratio=true,width=.49\unitlength]{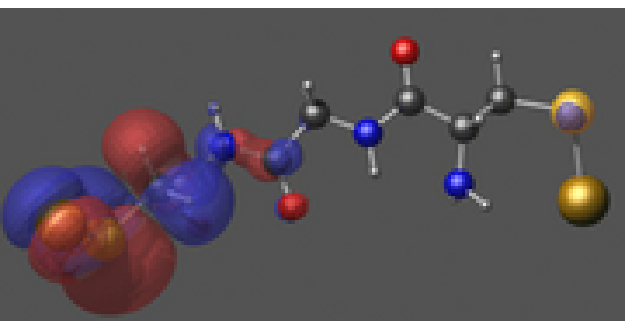}}
  \put(.35,.28){(b)}
  \put(.51,.26){\includegraphics[keepaspectratio=true,width=.49\unitlength]{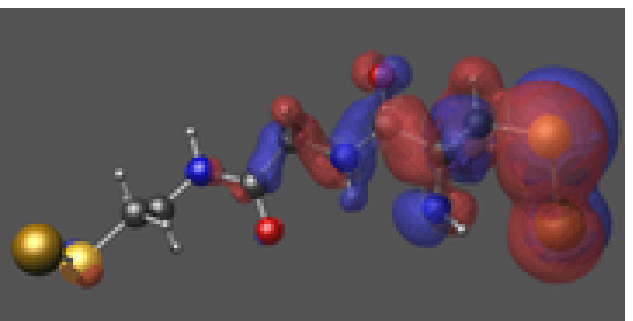}}
  \put(.85,.28){(c)}
  \put(0,0){\includegraphics[keepaspectratio=true,width=.49\unitlength]{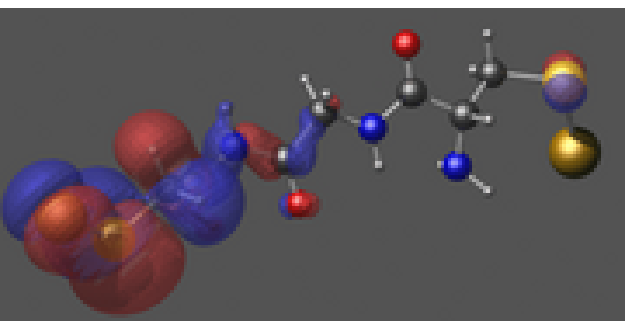}}
  \put(.35,.02){(d)}
  \put(.51,0){\includegraphics[keepaspectratio=true,width=.49\unitlength]{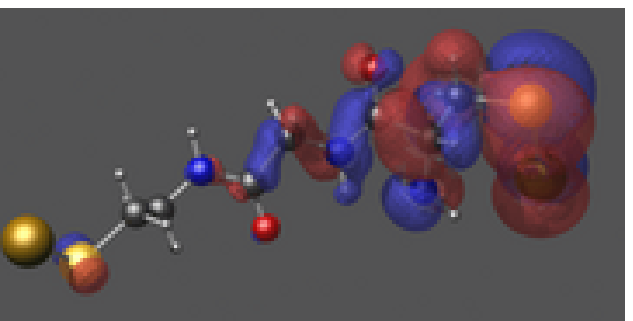}}
  \put(.85,.02){(e)}
\end{picture}
\caption{(Color online) (a) Equilibrium transmission functions $T(E,0)$ for
  cysteamine-Gly-Cys in the top--top configuration, for different distances
  between the two gold leads. Conductances of the metal--molecule--metal
  junction, shown in inset as a function of tip--tip distance, vary approximately
  linearly over a range of about 1\AA, beyond which no stable molecular structure bridges
  the two electrodes. Thus, the amount of stretching can affect conductance 
by a
  measurable amount, though not sufficiently to reconcile single-Au
  bonding geometries with the experimental results. For the hollow-hollow configuration, on the other
  hand, this range is much smaller, and so almost no variation in conductance
  is allowed. (b) and (c) show the $L$ and $R$ states, respectively, for the
  top-top simplified extended molecule with a tip-to-tip distance of
  14.2\AA. (d) and (e) show the same for a tip-to-tip distance of
  15.2\AA. Most of the stretching is accommodated by the bond angles at the
  gold-molecule interfaces, so that the principal change in (a) is peak widths.}
\label{stretching}
\end{figure}

Indeed, the tetrahedral $sp^3$ bonding of the hollow-bound
geometry is much less flexible, and its symmetry more restrictive,
than the linear $sp$ bond of the top-bound
geometry. Therefore, when the hollow-hollow geometry is stretched, the DFT geometry optimizations only allow
stretching of .2\AA\ or less (which affects $T(E)$ by 10\% or less) before the
molecular wire detaches from one electrode. By contrast,
we find stretching distances as large as 1\AA\ in 
top-bound geometries, and the conductance of unstretched top-bound
geometries can be as much as 3--4
times larger than that of the fully stretched configurations reported in Table
\ref{tab}, as shown in Fig.~\ref{stretching}. Since the
experimental data~\cite{xiaojacs04,xiaoac04} exhibit no strong stretching
dependence of conductance within a single plateau, we conclude that the
experimental results correspond to the
hollow-hollow geometry,\cite{ductilegold} in agreement with the
conclusion reached above based on the magnitude of
conductance.\cite{stretchingnote}

Kr{\"u}ger \emph{et al.}\ have suggested\cite{kruegerprl02} that when a
molecular wire thiol-bonded to gold contacts is stretched and eventually
breaks, the cleavage does not occur at the gold-thiol interface, but instead
that a gold atom, remaining attached to the thiol group, is pulled from
the contact. During this process, the system passes through several
intermediate, temporary configurations, including structures resembling the top-bound geometry
discussed above. While a complete analysis of this geometric evolution and
accompanying conductance changes is beyond the scope of the present work, we
note that the results shown in Table~\ref{tab} indicate that the transition
from hollow-hollow to a top-bound geometry at either end is associated with a
drop in conductance of more than an order of magnitude in each oligopeptide
molecule. Thus, in a stretching experiment, when the top-bound geometry forms,
the conductance plateau ends and the molecule appears to have simply separated
from one or both contacts.

Furthermore, during stretching, temporary intermediate configurations such as
those in Ref.~\onlinecite{kruegerprl02} may form, which in general are less
conducting than the original configuration. One probable intermediate
bonding geometry has the sulfur directly above a single gold atom embedded
\emph{within} the top atomic layer of the contact;
when one end of the molecule is bound in such a way to the gold, and the other
is in the hollow configuration, we find the molecular junction has a
conductance of 1/4 that of the hollow-hollow configuration. 
Indeed, weak features of lower conductance are seen in some
of the experimental conductance vs stretching curves of Ref.~\onlinecite{xiaojacs04}. They are
typically much shorter than the main plateau, and occur at its end,
consistent with configurations that are only
meta-stable and occasionally populated via the stretching mechanism.

To understand the transmission curves, it is helpful
to think in terms of a smaller extended molecule which includes only the two
to six gold atoms to which the sulfurs bond.
Transmission states near $E_F$ correspond to gold-sulfur interfacial orbitals of this
simplified extended molecule. Because there are two metal--molecule
interfaces, these states typically occur in pairs with similar energies. When the two
states are detuned from each other, they are largely localized at either end
of the molecule; when in resonance, they mix into bonding and
antibonding combinations (see Fig.~\ref{anticrossing}). For a symmetric
extended molecule at zero bias, the pairs
are always in resonance; oligopeptides, however, are intrinsically asymmetric,
so interfacial states
can be detuned from each other and localized at either end, even in
equilibrium.

Prominent peaks in the transmission spectra (Fig.~\ref{eq}) occur near the energies
of molecular orbitals which bridge the left and right leads. Thus, the 
resonances, and consequent hybridization, of levels otherwise localized on
opposite ends of the extended molecule play a dominant role in determining
electron transport. For example, in all molecules and bonding configurations
studied here, the two levels immediately below $E_F$ are paired interfacial
states, which we call $R$ (localized at the gold--molecule interface near the
amine terminus of the molecule) and $L$ (localized at the other
interface). The narrow transmission peaks associated with their bonding and
antibonding combinations are visible 1eV below $E_F$ in the double on-top
(dotted) curves of Fig.~\ref{eq}. Because the overlap of the gold $6s$ and
sulfur $3p$ orbital is strongly angle dependant,\cite{geometry} these
peaks each become broadened when their respective bonding geometries are changed
to hollow. Furthermore, the hollow-bonded versions of $R$ are more sensitive to the asymmetric
features of the oligopeptide molecules, resulting in a
significant detuning from $L$ (shown by the crosses in
Fig.~\ref{eq}). The transmission peak associated with these two states is thus
also severely broadened, becoming a smooth 
shoulder in $T(E,0)$.

\subsection{Current Rectification}

The movement and change under bias of $R$ and $L$, depicted in Fig.~\ref{anticrossing}, give
rise to the rectification observed by Xiao \emph{et al.}\ in
Ref.~\onlinecite{xiaoac04}. Positive bias raises (lowers) the
electrochemical potential of atomic orbitals on the amine (non-amine) side of
the molecule; consequently, under negative bias the energies of $R$ and $L$ 
move closer together. Near resonance, $R$ and $L$ hybridize into whole-molecule
states, allowing significant charge transport.

\begin{figure}
\setlength{\unitlength}{.5\columnwidth}
\begin{picture}(2,1)
  \put(0,0){\includegraphics[keepaspectratio=false,width=\columnwidth,height=.5\columnwidth]{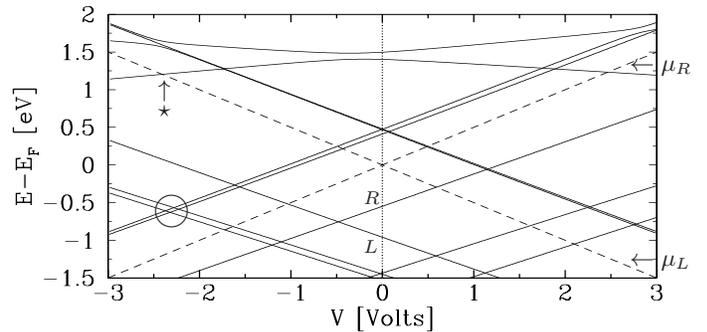}}
  \put(1.08,.39){\scriptsize{$R$}}
  \put(1.08,.24){\scriptsize{$L$}}
  \put(.49,.37){\circle{.1}}
  \put(1.9,.8){$\leftarrow\mu_R$}
  \put(1.9,.2){$\leftarrow\mu_L$}
  \put(.45,.72){$\uparrow$}
  \put(.45,.66){$\star$}
\end{picture}
\caption{Movement with applied bias of the generalized eigenvalues\cite{amountstretch} of the six-gold-atom simplified extended
  molecule for cysteamine-Gly-Cys in the hollow-hollow configuration. Some
  lines are doubled by degrees of freedom in the gold contacts; when more
  gold atoms are included, the molecular orbitals split further, eventually
  developing into
  the continuous density of states of an infinite system. Note
  that the apparent intersections are actually avoided crossings. In
  equilibrium, states $R$ and $L$ are detuned by .4eV due to molecular asymmetry. They come into resonance
  only under negative bias, causing current rectification. 
  Without these two states, the location of anticrossings, and thus the $I$--$V$
  characteristic, would be largely symmetric. The dashed lines represent the
  electrochemical potentials of the right and left electrodes, $\mu_R$ and $\mu_L$. NDR is the results of
  increasing bias beyond the last anticrossing in a group; an example is circled.}
\label{levels}
\end{figure}

Figure \ref{levels} shows the movement under bias of several generalized
eigenenergies of the simplified extended molecule for hollow-hollow bonded
cysteamine-Gly-Cys.\cite{amountstretch} All of the states in the region around $E_F$ occur in
right/left pairs like $R$ and $L$, but only $R$ and $L$ are detuned from each
other in equilibrium. Without $R$ and $L$, the
locations of anticrossings are nearly symmetric about
$V=0$. Rectification is 
a result of the splitting between
these two states, itself a direct consequence of oligopeptide asymmetry.

\begin{figure}
\setlength{\unitlength}{.5\columnwidth}
\begin{picture}(2,2)
\put(1.25,1.18){\includegraphics[keepaspectratio=true,width=.6\unitlength]{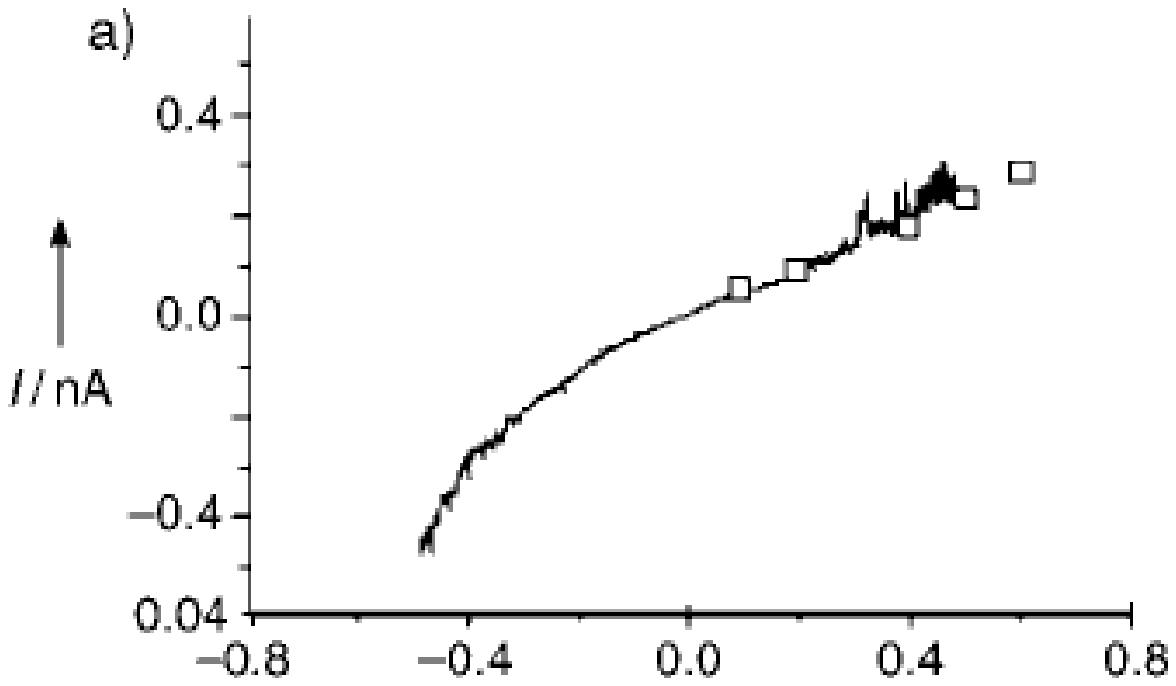}}
\put(1.3,1.5){\color{white}\circle*{.1}}
\put(.2,.18){\includegraphics[keepaspectratio=true,width=.66\unitlength]{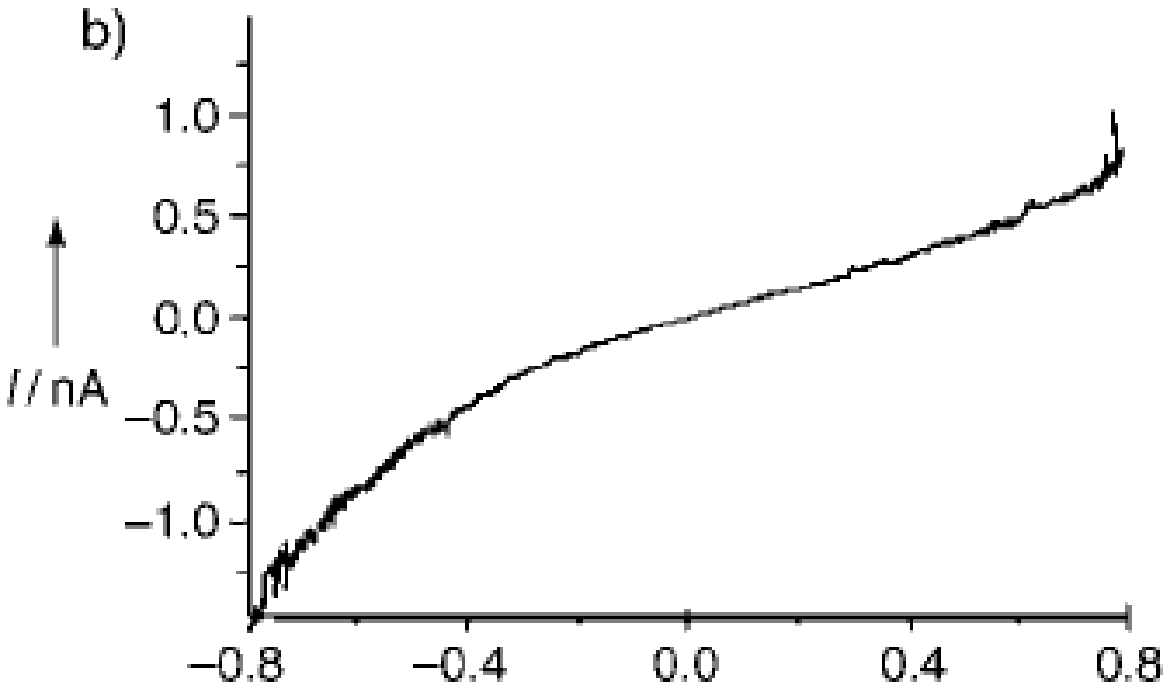}}
\put(.25,.55){\color{white}\circle*{.08}}
\put(1.3,.18){\includegraphics[keepaspectratio=true,width=.6\unitlength]{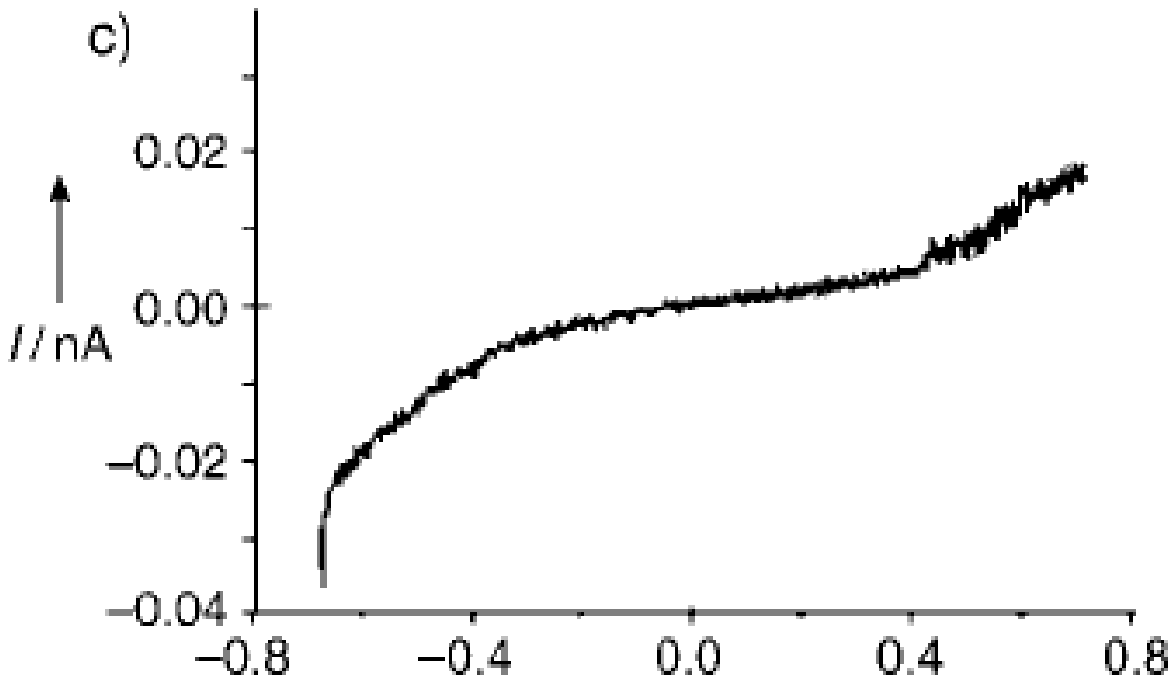}}
\put(1.35,.5){\color{white}\circle*{.08}}
\put(1.4,.5){\color{white}\circle*{.08}}
\put(0,1){\includegraphics[keepaspectratio=true,width=\unitlength]{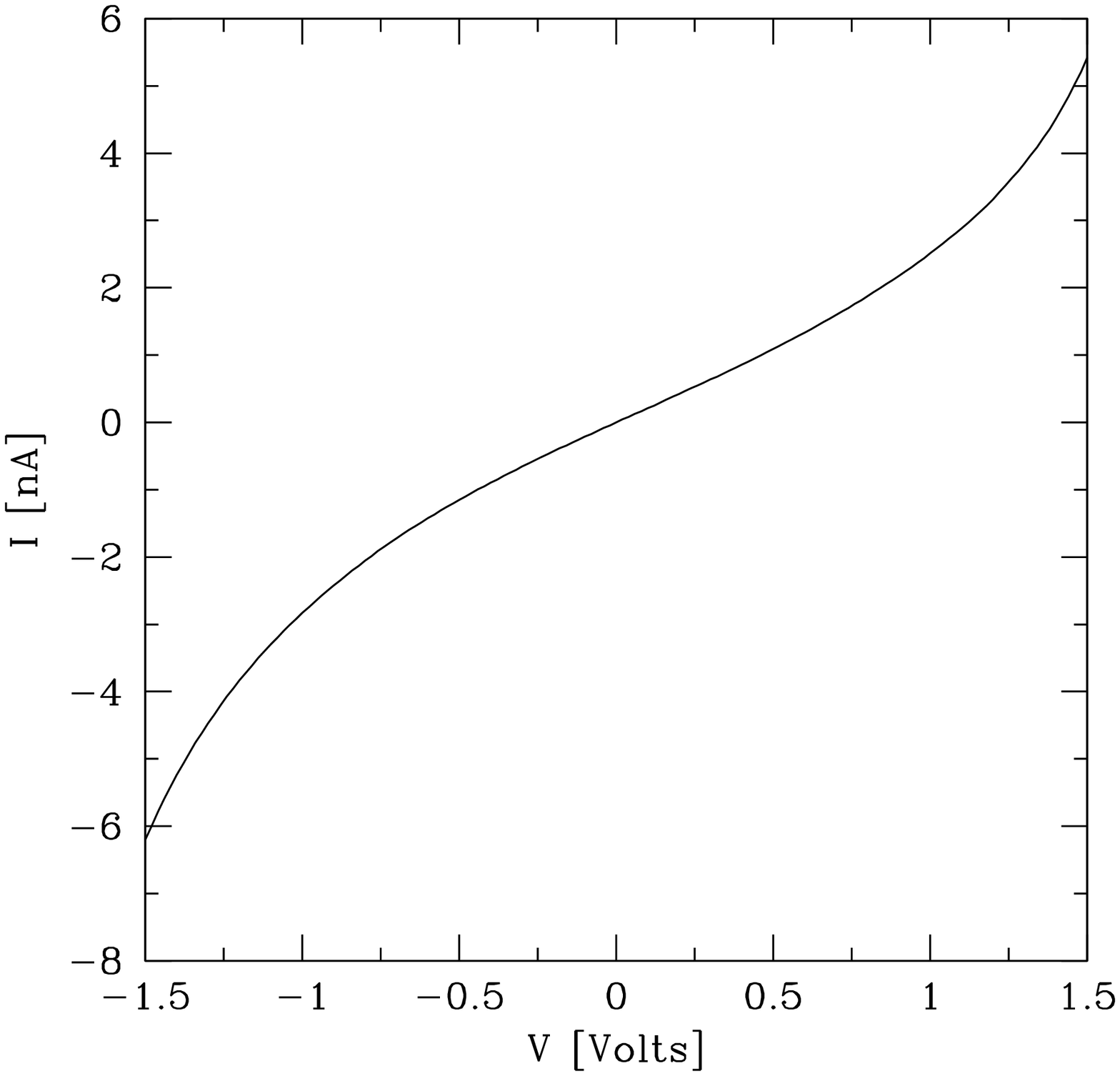}}
\put(1,1){\includegraphics[keepaspectratio=true,width=\unitlength]{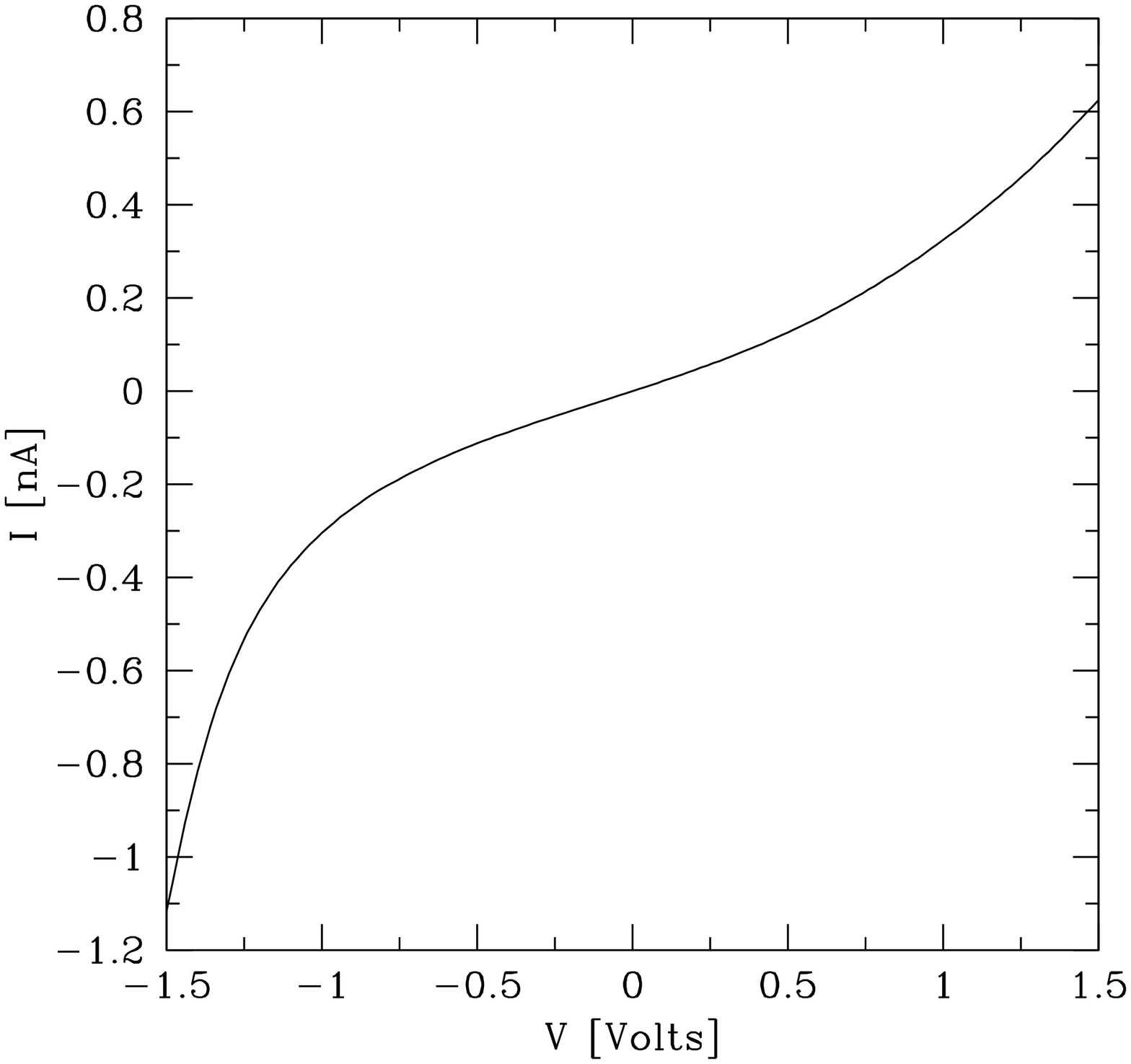}}
\put(0,0){\includegraphics[keepaspectratio=true,width=\unitlength]{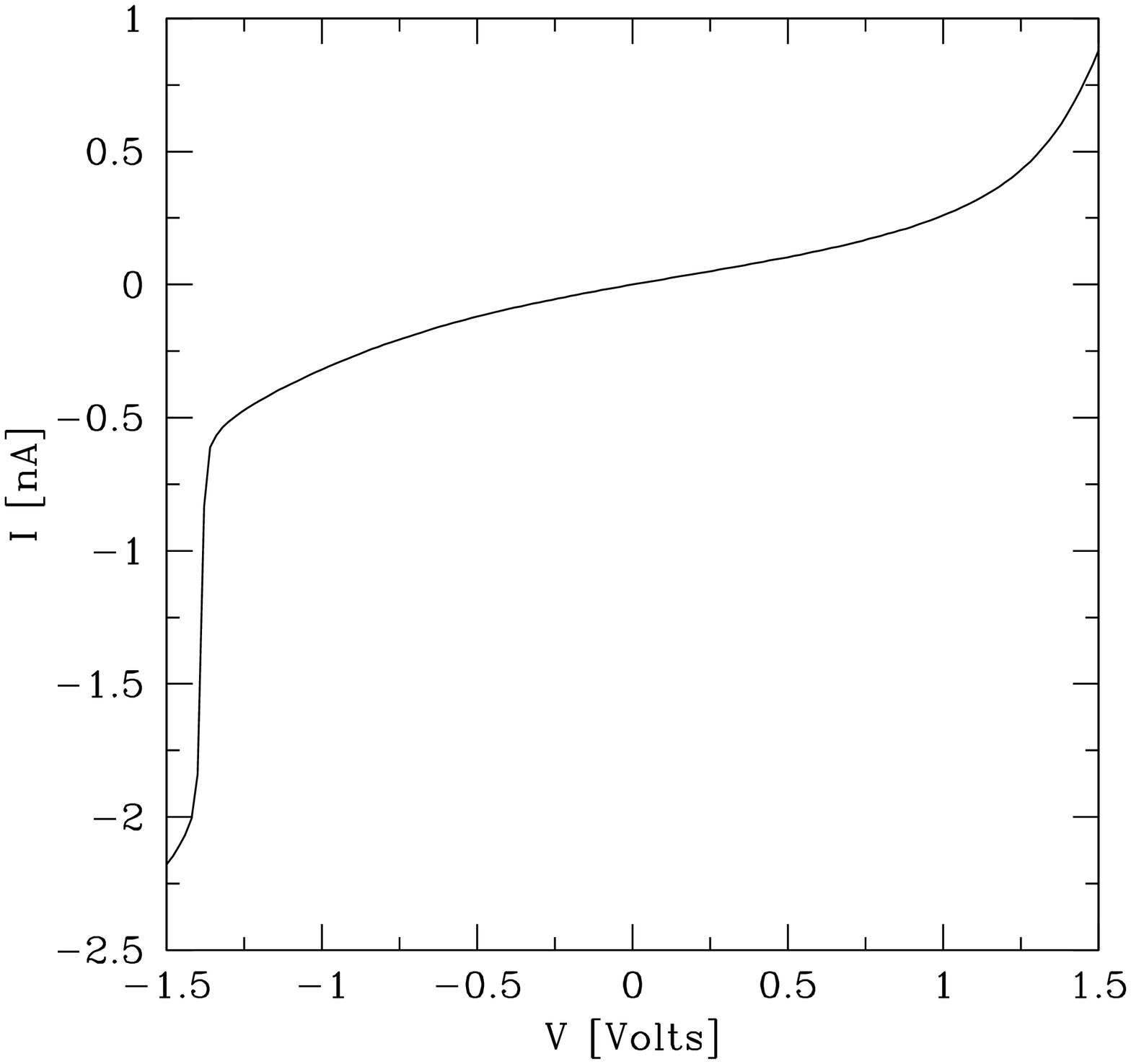}}
\put(1,0){\includegraphics[keepaspectratio=true,width=\unitlength]{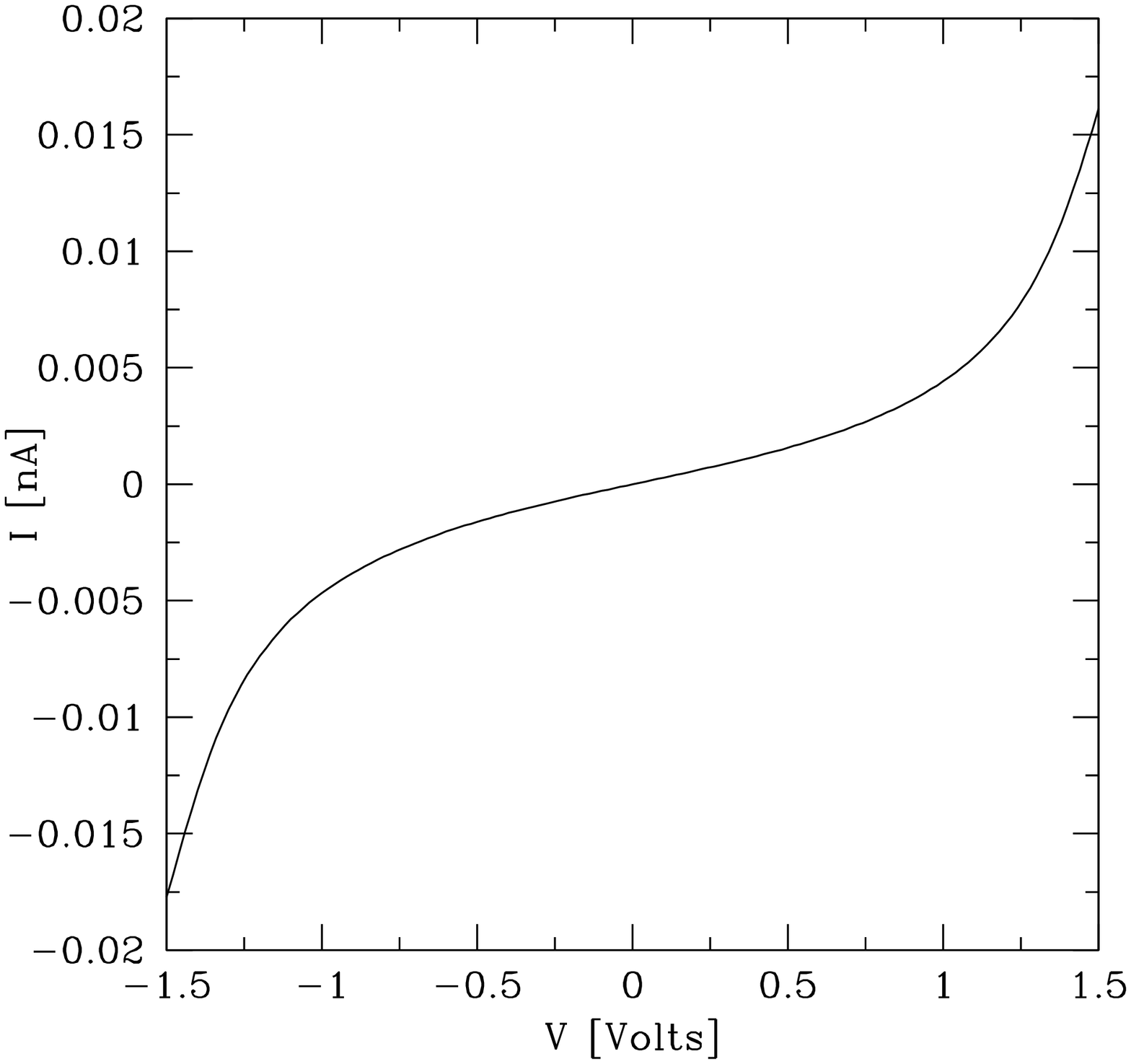}}
\put(.18,1.8){cysteamine-Cys}
\put(1.17,1.8){cysteamine-Gly-Cys}
\put(.18,.8){Cys-Gly-Cys}
\put(1.17,.8){\footnotesize{cysteamine-Gly-Gly-Cys}}
\end{picture}
\caption{Calculated current-voltage characteristics of four single-molecule
  oligopeptide devices.\cite{amountstretch} The scale and asymmetry of the current are comparable
  to those found in the 
  experimental studies of Ref.~\onlinecite{xiaoac04}, reproduced here in
  inset.}
\label{IV}
\end{figure}

Figure~\ref{IV} gives the calculated current-voltage characteristic for each
molecule,\cite{amountstretch} showing this rectification.
While the resonance depicted in Fig.~\ref{anticrossing} is comparatively
narrow and outside the energy window between $\mu_L$ and $\mu_R$, rectification extends beyond 2V in the
full systems due to the strong hollow-site broadening of $R$ and $L$.
For the three molecules for
which Xiao \emph{et al.} made $I$--$V$ measurements (shown in inset), the
calculated rectification is in excellent qualitative agreement with experimental
data. 
We find a current ratio of about
150\% for cysteamine-Gly-Cys and Cys-Gly-Cys, similar to Ref.~\onlinecite{xiaoac04}.
Cysteamine-Gly-Gly-Cys exhibits less rectification,
both in our model and the experimental data: This is due to its much smaller
detuning between $R$ and $L$ at zero bias.

\subsection{Negative Differential Resistance}

Negative differential resistance ($dI/dV<0$) can result from applied bias destroying a resonance between
transmitting interfacial states.\cite{dalgleishnl06,dalgleishprb06} As the
bias increases past an anticrossing in the metal--oligopeptide--metal systems,
the two interfacial states are taken further apart in energy. They therefore
mix less well, and become localized at the individual metal--molecule
interfaces. As the bias voltage increases, then, electron transport through
these states ``shuts off'', creating the potential for NDR.

\begin{figure}
\setlength{\unitlength}{.5\columnwidth}
\begin{picture}(2,1)
\put(0,0){\includegraphics[keepaspectratio=false,height=\unitlength,width=2\unitlength]{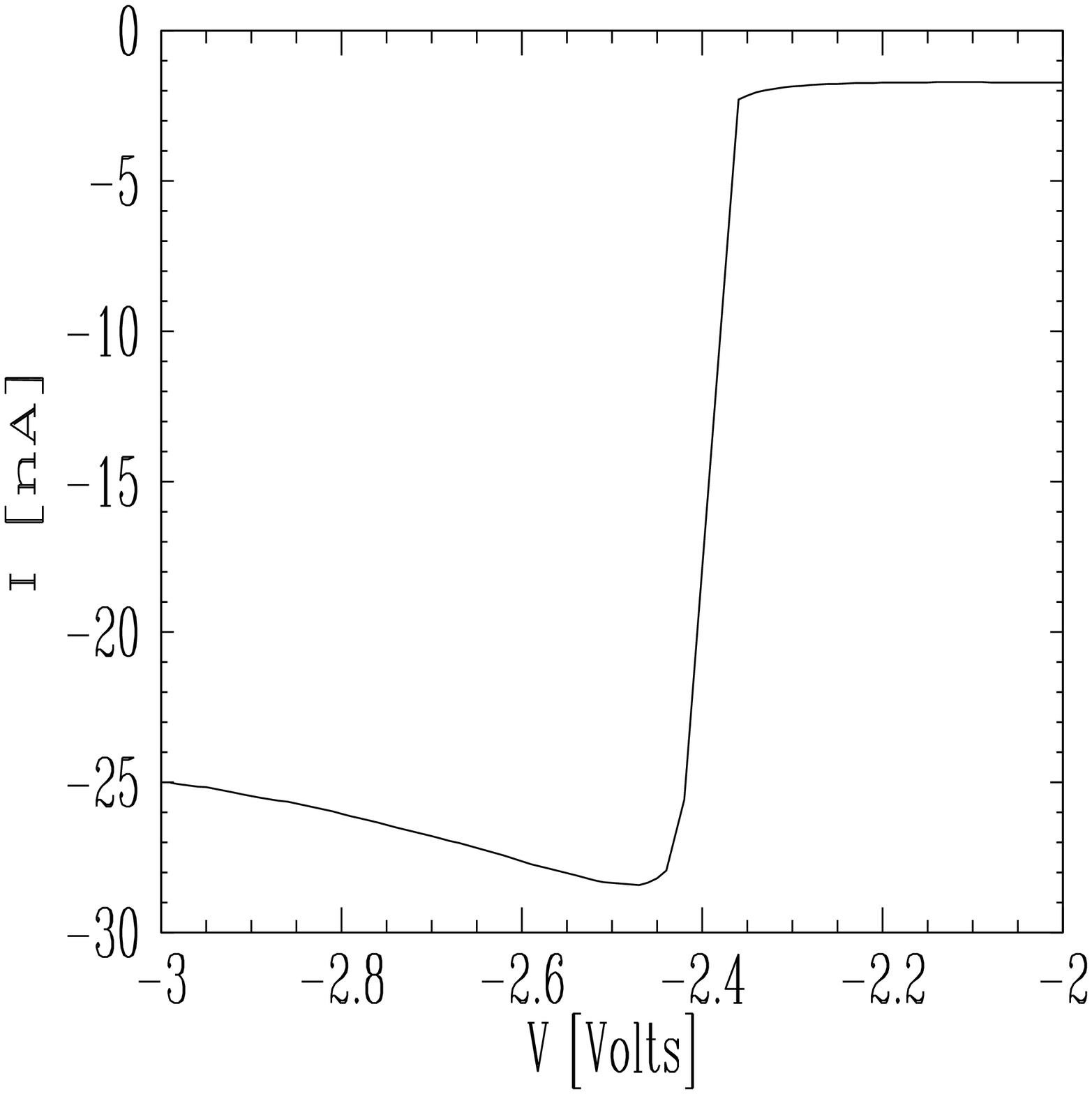}}
\put(.295,.41){\includegraphics[keepaspectratio=false,height=.5\unitlength,width=1\unitlength]{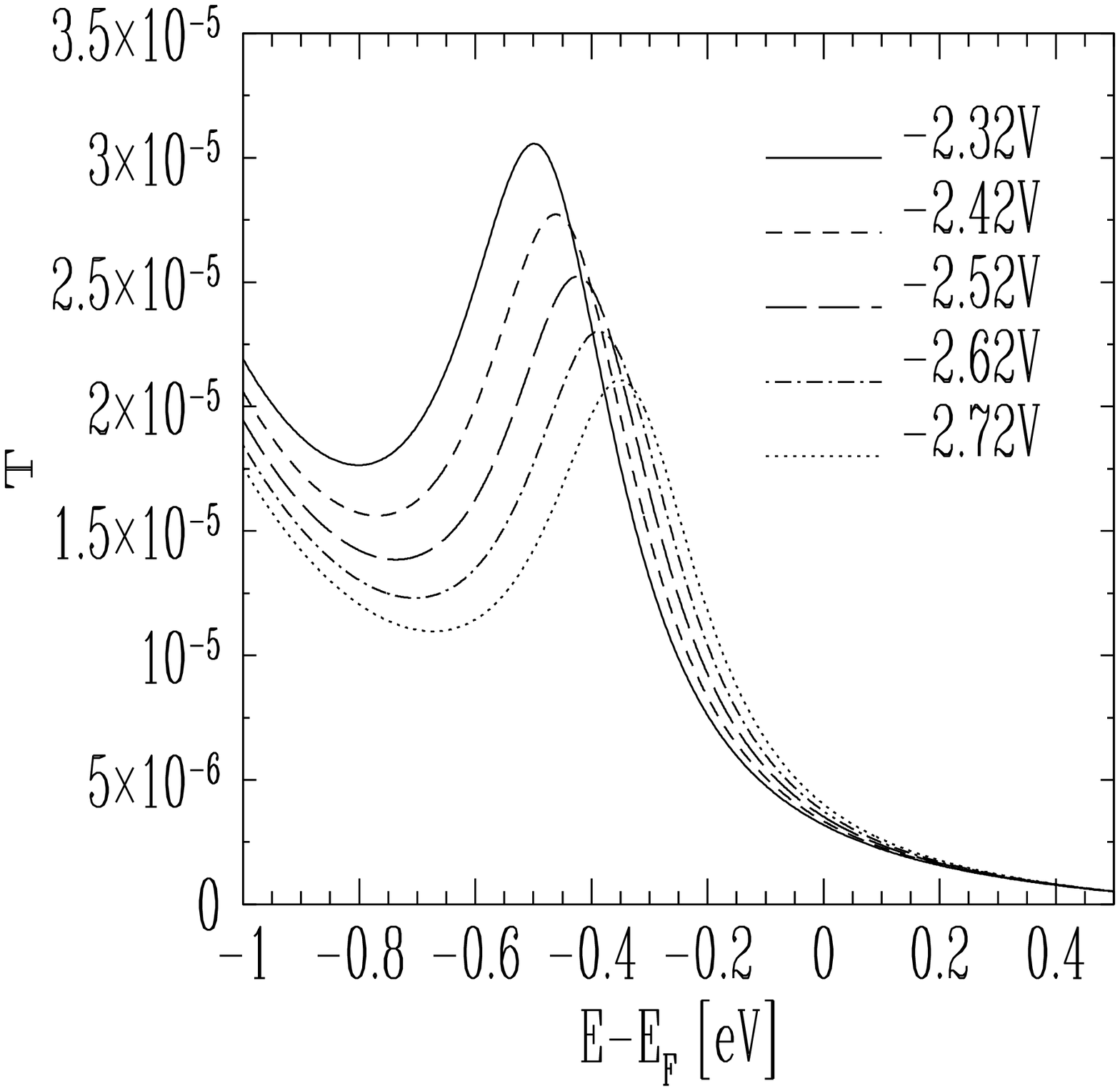}}
\end{picture}
\caption{$I$--$V$ characteristic for the hollow-hollow gold--cysteamine-Gly-Cys--gold system\cite{amountstretch} in
the medium-bias regime, showing NDR. Other oligopeptide molecules have
qualitatively similar characteristics. (inset) Corresponding $T(E,V)$ curves.
NDR corresponds to a
gradual decrease in the transmission peak's strength as voltage becomes more negative,
due to departure from a resonant crossing of interfacial states (the circle in Fig.~\ref{levels}).}
\label{NDR}
\end{figure}

Such effects are masked while the current is still small: The
dominant effect is the increasing current from a series of crossings and states
entering the window of integration.
Nevertheless, once the voltage is large enough that an entire
neighborhood of interfacial resonances has been encompassed, further applied
bias
yields NDR. 
One such crossing is indicated by the circle in Fig.~\ref{levels}; its $I$--$V$ characteristic
and relevant transmission properties are shown in Fig.~\ref{NDR}. As $V$ becomes more negative, the peak in $T(E)$ associated with the nearly degenerate hybridized
states diminishes. After the final current step, due to a mid-molecule orbital
($\star$ in Fig.~\ref{levels})
entering the window of integration, NDR results.
The bias voltages required to observe this important phenomenon are
only moderately larger than those already achieved
experimentally.\cite{xiaoac04}

The magnitude of NDR is determined by the participating levels' electrostatic and
quantum mechanical couplings to the leads. While the first is
necessarily high for interfacial levels (hence the saturated slope of most
lines in Fig.~\ref{levels}), the quantum mechanical coupling, or broadening, of
an interfacial level depends mainly on the bond geometry at the metal--molecule
interfaces. For hollow bonding, for example, interfacial states are very well
broadened, as can be seen from the hollow-hollow (solid) curves of Fig.~\ref{eq}. On the
other hand, if one or both of the interfaces is in the top-bound geometry, the
related NDR is stronger.

Like their
rectifying properties, the NDR that we predict in oligopeptides
is a direct result of the dominance of resonant crossings of 
interfacial states in charge transport. Since this is a property of all thiol-
(\emph{i.e.} cysteamine- or cysteine-) bonded oligopeptide molecules, both
effects are robust, and observable in the entire class of
such molecules. Because of the limitless customizability and simple
fabrication of oligopeptides, the effects can be tailored to meet a wide variety of device
needs.

\section{\label{sec:conclusion}Conclusions}

In summary, we have developed a theoretical model which explains all of the
experimental data on charge transport through oligopeptide
molecules. We obtain quantitative agreement with experiment
by using a combination of appropriate, standard theoretical models 
(DFT for the molecular geometry;
the linearized Debye--H\"uckel model for the potential profile, including the
electrostatic screening effect of the electrolyte solution;
 and the extended H\"uckel model of Hoffman and co-workers\cite{hoffmanjcp63,ammeterjacs78} for 
the electronic structure) 
and careful consideration of the experimental systems. 
We use no fitting of any kind to any experimental data involving electrical
conduction through molecules.
Quantitative agreement without the use of fitting parameters allows us to extract
detailed chemical and physical information from the experiments, such as the
conclusion that both ends of the molecule are bound to hollow sites on the
Au(111) surfaces.

We understand the observed current rectification of oligopeptides as due to
a hybridization of interfacial states lying at either end of the
molecule, naturally detuned from each other in equilibrium due to the
intrinsic asymmetry of peptide chains. When bias is applied in one direction,
these states are brought into resonance with each other, forming two
whole-molecule states which allow a stronger current to flow; such a resonance
does not occur when the molecule is biased in the opposite direction. Our
calculated current rectifying ratios agree well with those measured in
experiment, again without the use of fitting parameters.

Finally, our model predicts NDR will appear at moderately higher bias
voltages than have been explored experimentally  in Refs.~\onlinecite{xiaojacs04} and
\onlinecite{xiaoac04}. This phenomenon is understood as
the natural counterpart to the rectifying physics: As bias is increased past
the anticrossing of the two interfacial states, the states once again become
localized at the metal-molecule interfaces, and current flow through these
channels decreases. Since NDR has many attendant device applications, and
oligopeptides offer a unique combination of customizability and scalability,
it is our hope that this work will inspire much further study of their
electron transport properties.

\vspace*{2em}
\begin{acknowledgments}
We thank Nongjian Tao, Joshua Hihath, and Ross Hill for useful
discussions. This work was supported by NSERC and the Canadian Institute for
Advanced Research. Some numerical computations presented in this work were
performed on WestGrid computing resources, which are funded in part by the
Canada Foundation for Innovation, Alberta Innovation and Science, BC Advanced
Education, and the participating research institutions. WestGrid equipment is
provided by IBM, Hewlett Packard, and SGI.
\end{acknowledgments}

\bibliography{oligopeptides.bib}

\begin{thebibliography}{32}
\expandafter\ifx\csname natexlab\endcsname\relax\def\natexlab#1{#1}\fi
\expandafter\ifx\csname bibnamefont\endcsname\relax
  \def\bibnamefont#1{#1}\fi
\expandafter\ifx\csname bibfnamefont\endcsname\relax
  \def\bibfnamefont#1{#1}\fi
\expandafter\ifx\csname citenamefont\endcsname\relax
  \def\citenamefont#1{#1}\fi
\expandafter\ifx\csname url\endcsname\relax
  \def\url#1{\texttt{#1}}\fi
\expandafter\ifx\csname urlprefix\endcsname\relax\def\urlprefix{URL }\fi
\providecommand{\bibinfo}[2]{#2}
\providecommand{\eprint}[2][]{\url{#2}}

\bibitem[{\citenamefont{Gilmanshin}(1993)}]{gilmanshin93}
\bibinfo{author}{\bibfnamefont{R.~I.} \bibnamefont{Gilmanshin}},
  \emph{\bibinfo{title}{\emph{In K. Sienicki, ed.} Molecular Electronics and
  Molecular Electronic Devices, Vol. 2}} (\bibinfo{publisher}{CRC Press},
  \bibinfo{address}{Boca Raton}, \bibinfo{year}{1993}), pp.
  \bibinfo{pages}{1--78}.

\bibitem[{\citenamefont{Xiao et~al.}(2004{\natexlab{a}})\citenamefont{Xiao, Xu,
  and Tao}}]{xiaojacs04}
\bibinfo{author}{\bibfnamefont{X.}~\bibnamefont{Xiao}},
  \bibinfo{author}{\bibfnamefont{B.}~\bibnamefont{Xu}}, \bibnamefont{and}
  \bibinfo{author}{\bibfnamefont{N.}~\bibnamefont{Tao}}, \bibinfo{journal}{J.
  Am.\ Chem.\ Soc.} \textbf{\bibinfo{volume}{126}}, \bibinfo{pages}{5370}
  (\bibinfo{year}{2004}{\natexlab{a}}).

\bibitem[{\citenamefont{Xiao et~al.}(2004{\natexlab{b}})\citenamefont{Xiao, Xu,
  and Tao}}]{xiaoac04}
\bibinfo{author}{\bibfnamefont{X.}~\bibnamefont{Xiao}},
  \bibinfo{author}{\bibfnamefont{B.}~\bibnamefont{Xu}}, \bibnamefont{and}
  \bibinfo{author}{\bibfnamefont{N.}~\bibnamefont{Tao}},
  \bibinfo{journal}{Angew. Chem. Int. Ed.} \textbf{\bibinfo{volume}{43}},
  \bibinfo{pages}{6148} (\bibinfo{year}{2004}{\natexlab{b}}).

\bibitem[{\citenamefont{Xu and Tao}(2003)}]{xuscience03}
\bibinfo{author}{\bibfnamefont{B.}~\bibnamefont{Xu}} \bibnamefont{and}
  \bibinfo{author}{\bibfnamefont{N.~J.} \bibnamefont{Tao}},
  \bibinfo{journal}{Science} \textbf{\bibinfo{volume}{301}},
  \bibinfo{pages}{1221} (\bibinfo{year}{2003}).

\bibitem[{\citenamefont{Xu et~al.}(2003)\citenamefont{Xu, Xiao, and
  Tao}}]{xujacs03}
\bibinfo{author}{\bibfnamefont{B.}~\bibnamefont{Xu}},
  \bibinfo{author}{\bibfnamefont{X.}~\bibnamefont{Xiao}}, \bibnamefont{and}
  \bibinfo{author}{\bibfnamefont{N.~J.} \bibnamefont{Tao}},
  \bibinfo{journal}{J. Am.\ Chem.\ Soc.} \textbf{\bibinfo{volume}{125}},
  \bibinfo{pages}{16164} (\bibinfo{year}{2003}).

\bibitem[{\citenamefont{Xiao et~al.}(2004{\natexlab{c}})\citenamefont{Xiao, Xu,
  and Tao}}]{xiaonl04}
\bibinfo{author}{\bibfnamefont{X.}~\bibnamefont{Xiao}},
  \bibinfo{author}{\bibfnamefont{B.}~\bibnamefont{Xu}}, \bibnamefont{and}
  \bibinfo{author}{\bibfnamefont{N.~J.} \bibnamefont{Tao}},
  \bibinfo{journal}{Nano Lett.} \textbf{\bibinfo{volume}{4}},
  \bibinfo{pages}{267} (\bibinfo{year}{2004}{\natexlab{c}}).

\bibitem[{\citenamefont{Datta}(2003)}]{datta}
\bibinfo{author}{\bibfnamefont{S.}~\bibnamefont{Datta}},
  \emph{\bibinfo{title}{Electron Transport in Mesoscopic Systems}},
  vol.~\bibinfo{volume}{3} of \emph{\bibinfo{series}{Cambridge Studies in
  Semiconductor Physics and Microelectronic Engineering}}
  (\bibinfo{publisher}{Cambridge University Press},
  \bibinfo{address}{Cambridge, UK}, \bibinfo{year}{2003}), ISBN
  \bibinfo{isbn}{0 521 59943 1}.

\bibitem{amountstretch}
\bibinfo{nonte}{All of the numerical results presented in this article (with
  the exception of those in Fig.~\ref{stretching}) are for the largest value of
  the spacing between the gold electrodes (for each molecule and bond geometry
  between the molecule and gold contacts) for which the \emph{ab initio}
  density functional theory geometry optimization converges to a geometry in
  which the molecule bridges the two leads. This is consistent with the
  experimental procedures of Refs.~\onlinecite{xiaojacs04,xiaoac04,xuscience03,xujacs03,xiaonl04}, in which the
  transport measurements were performed on stretched metal--molecule--metal junctions.}

\bibitem[{\citenamefont{van~der Ziel}(1983)}]{vanderzielsse83}
\bibinfo{author}{\bibfnamefont{A.}~\bibnamefont{van~der Ziel}},
  \bibinfo{journal}{Solid-State Electron.} \textbf{\bibinfo{volume}{26}},
  \bibinfo{pages}{333} (\bibinfo{year}{1983}).

\bibitem[{\citenamefont{Brown et~al.}(1991)\citenamefont{Brown,
  S{\"o}derstr{\"o}m, Parker, Mahoney, Molvar, and McGill}}]{brownapl91}
\bibinfo{author}{\bibfnamefont{E.~R.} \bibnamefont{Brown}},
  \bibinfo{author}{\bibfnamefont{J.~R.} \bibnamefont{S{\"o}derstr{\"o}m}},
  \bibinfo{author}{\bibfnamefont{C.~D.} \bibnamefont{Parker}},
  \bibinfo{author}{\bibfnamefont{L.~J.} \bibnamefont{Mahoney}},
  \bibinfo{author}{\bibfnamefont{K.~M.} \bibnamefont{Molvar}},
  \bibnamefont{and} \bibinfo{author}{\bibfnamefont{T.~C.}
  \bibnamefont{McGill}}, \bibinfo{journal}{Appl.\ Phys.\ Lett.}
  \textbf{\bibinfo{volume}{58}}, \bibinfo{pages}{2291} (\bibinfo{year}{1991}).

\bibitem[{\citenamefont{Broekaert et~al.}(1998)\citenamefont{Broekaert, Brar,
  van~der Wagt, Seabaugh, Morris, Moise, Beam, and
  Frazier}}]{broekaertieeejssc98}
\bibinfo{author}{\bibfnamefont{T.~P.~E.} \bibnamefont{Broekaert}},
  \bibinfo{author}{\bibfnamefont{B.}~\bibnamefont{Brar}},
  \bibinfo{author}{\bibfnamefont{J.~P.~A.} \bibnamefont{van~der Wagt}},
  \bibinfo{author}{\bibfnamefont{A.~C.} \bibnamefont{Seabaugh}},
  \bibinfo{author}{\bibfnamefont{F.~J.} \bibnamefont{Morris}},
  \bibinfo{author}{\bibfnamefont{T.~S.} \bibnamefont{Moise}},
  \bibinfo{author}{\bibfnamefont{E.~A.} \bibnamefont{Beam},
  \bibfnamefont{III}}, \bibnamefont{and} \bibinfo{author}{\bibfnamefont{G.~A.}
  \bibnamefont{Frazier}}, \bibinfo{journal}{IEEE J. Solid-State Circuits}
  \textbf{\bibinfo{volume}{33}}, \bibinfo{pages}{1342} (\bibinfo{year}{1998}).

\bibitem[{\citenamefont{Mathews et~al.}(1999)\citenamefont{Mathews, Sage,
  Sollner, Calawa, Chen, Mahoney, Maki, and Molvar}}]{mathewspieee99}
\bibinfo{author}{\bibfnamefont{R.~H.} \bibnamefont{Mathews}},
  \bibinfo{author}{\bibfnamefont{J.~P.} \bibnamefont{Sage}},
  \bibinfo{author}{\bibfnamefont{T.~C. L.~G.} \bibnamefont{Sollner}},
  \bibinfo{author}{\bibfnamefont{S.~D.} \bibnamefont{Calawa}},
  \bibinfo{author}{\bibfnamefont{C.-L.} \bibnamefont{Chen}},
  \bibinfo{author}{\bibfnamefont{L.~J.} \bibnamefont{Mahoney}},
  \bibinfo{author}{\bibfnamefont{P.~A.} \bibnamefont{Maki}}, \bibnamefont{and}
  \bibinfo{author}{\bibfnamefont{K.~M.} \bibnamefont{Molvar}},
  \bibinfo{journal}{Proc.\ IEEE} \textbf{\bibinfo{volume}{87}},
  \bibinfo{pages}{596} (\bibinfo{year}{1999}).

\bibitem[{gau()}]{gaussian}
\bibinfo{note}{Gaussian 98, Revision A.11.3, M. J. Frisch \emph{et al.}.}

\bibitem{broadeningnote}
\bibinfo{note}{Inclusion of such a large metallic nanocluster between leads
  and molecule allows for a quantitative treatment of peak broadenings, which
  are essential to the phenomena we discuss in this work. We
  have verified that the clusters we use are large enough that increasing
  their size further does not significantly affect the results.}

\bibitem[{\citenamefont{Datta et~al.}(1997)\citenamefont{Datta, Tian, Hong,
  Reifenberger, Henderson, and Kubiak}}]{dattaprl97}
\bibinfo{author}{\bibfnamefont{S.}~\bibnamefont{Datta}},
  \bibinfo{author}{\bibfnamefont{W.}~\bibnamefont{Tian}},
  \bibinfo{author}{\bibfnamefont{S.}~\bibnamefont{Hong}},
  \bibinfo{author}{\bibfnamefont{R.}~\bibnamefont{Reifenberger}},
  \bibinfo{author}{\bibfnamefont{J.~I.} \bibnamefont{Henderson}},
  \bibnamefont{and} \bibinfo{author}{\bibfnamefont{C.~P.}
  \bibnamefont{Kubiak}}, \bibinfo{journal}{Phys.\ Rev.\ Lett.}
  \textbf{\bibinfo{volume}{79}}, \bibinfo{pages}{2530} (\bibinfo{year}{1997}).

\bibitem[{\citenamefont{Emberly and Kirczenow}(2001)}]{emberlyprb01}
\bibinfo{author}{\bibfnamefont{E.~G.} \bibnamefont{Emberly}} \bibnamefont{and}
  \bibinfo{author}{\bibfnamefont{G.}~\bibnamefont{Kirczenow}},
  \bibinfo{journal}{Phys.\ Rev.\ B} \textbf{\bibinfo{volume}{64}},
  \bibinfo{pages}{235412} (\bibinfo{year}{2001}).

\bibitem[{\citenamefont{Reed et~al.}(1997)\citenamefont{Reed, Zhou, Muller,
  Burgin, and Tour}}]{reedscience97}
\bibinfo{author}{\bibfnamefont{M.~A.} \bibnamefont{Reed}},
  \bibinfo{author}{\bibfnamefont{C.}~\bibnamefont{Zhou}},
  \bibinfo{author}{\bibfnamefont{C.~J.} \bibnamefont{Muller}},
  \bibinfo{author}{\bibfnamefont{T.~P.} \bibnamefont{Burgin}},
  \bibnamefont{and} \bibinfo{author}{\bibfnamefont{J.~M.} \bibnamefont{Tour}},
  \bibinfo{journal}{Science} \textbf{\bibinfo{volume}{278}},
  \bibinfo{pages}{252} (\bibinfo{year}{1997}).

\bibitem[{\citenamefont{Kushmerick et~al.}(2002)\citenamefont{Kushmerick, Holt,
  Yang, Naciri, Moore, and Shashidhar}}]{kushmerickprl02}
\bibinfo{author}{\bibfnamefont{J.~G.} \bibnamefont{Kushmerick}},
  \bibinfo{author}{\bibfnamefont{D.~B.} \bibnamefont{Holt}},
  \bibinfo{author}{\bibfnamefont{J.~C.} \bibnamefont{Yang}},
  \bibinfo{author}{\bibfnamefont{J.}~\bibnamefont{Naciri}},
  \bibinfo{author}{\bibfnamefont{M.~H.} \bibnamefont{Moore}}, \bibnamefont{and}
  \bibinfo{author}{\bibfnamefont{R.}~\bibnamefont{Shashidhar}},
  \bibinfo{journal}{Phys.\ Rev.\ Lett.} \textbf{\bibinfo{volume}{89}},
  \bibinfo{pages}{086802} (\bibinfo{year}{2002}).

\bibitem[{\citenamefont{Hoffman}(1963)}]{hoffmanjcp63}
\bibinfo{author}{\bibfnamefont{R.}~\bibnamefont{Hoffman}}, \bibinfo{journal}{J.
  Chem.\ Phys.} \textbf{\bibinfo{volume}{39}}, \bibinfo{pages}{1397}
  (\bibinfo{year}{1963}).

\bibitem[{\citenamefont{Ammeter et~al.}(1978)\citenamefont{Ammeter, B{\"u}rgi,
  Thibeault, and Hoffman}}]{ammeterjacs78}
\bibinfo{author}{\bibfnamefont{J.~H.} \bibnamefont{Ammeter}},
  \bibinfo{author}{\bibfnamefont{H.-B.} \bibnamefont{B{\"u}rgi}},
  \bibinfo{author}{\bibfnamefont{J.~C.} \bibnamefont{Thibeault}},
  \bibnamefont{and} \bibinfo{author}{\bibfnamefont{R.}~\bibnamefont{Hoffman}},
  \bibinfo{journal}{J. Am.\ Chem.\ Soc.} \textbf{\bibinfo{volume}{100}},
  \bibinfo{pages}{3686} (\bibinfo{year}{1978}).

\bibitem[{\citenamefont{Cerd{\'a} and Soria}(2000)}]{cerdaprb00}
\bibinfo{author}{\bibfnamefont{J.}~\bibnamefont{Cerd{\'a}}} \bibnamefont{and}
  \bibinfo{author}{\bibfnamefont{F.}~\bibnamefont{Soria}},
  \bibinfo{journal}{Phys.\ Rev.\ B} \textbf{\bibinfo{volume}{61}},
  \bibinfo{pages}{7965} (\bibinfo{year}{2000}).

\bibitem[{\citenamefont{Kirczenow et~al.}(2005)\citenamefont{Kirczenow, Piva,
  and Wolkow}}]{kirczenowprb05}
\bibinfo{author}{\bibfnamefont{G.}~\bibnamefont{Kirczenow}},
  \bibinfo{author}{\bibfnamefont{P.~G.} \bibnamefont{Piva}}, \bibnamefont{and}
  \bibinfo{author}{\bibfnamefont{R.~A.} \bibnamefont{Wolkow}},
  \bibinfo{journal}{Phys.\ Rev.\ B} \textbf{\bibinfo{volume}{72}},
  \bibinfo{pages}{245306} (\bibinfo{year}{2005}).

\bibitem[{\citenamefont{Kienle et~al.}(2006)\citenamefont{Kienle, Cerda, and
  Ghosh}}]{kienlejap06}
\bibinfo{author}{\bibfnamefont{D.}~\bibnamefont{Kienle}},
  \bibinfo{author}{\bibfnamefont{J.~I.} \bibnamefont{Cerda}}, \bibnamefont{and}
  \bibinfo{author}{\bibfnamefont{A.~W.} \bibnamefont{Ghosh}},
  \bibinfo{journal}{J. Appl.\ Phys.} \textbf{\bibinfo{volume}{100}},
  \bibinfo{pages}{043714} (\bibinfo{year}{2006}).

\bibitem[{bin()}]{bind}
\bibinfo{note}{The numerical implementation used was the {\tt{YAeHMOP}} package
  (ver.\ 3.0), by G. A. Landrum and W. V. Glassey, with the default extended
  H\"uckel perscription for that program, given in Ref.~\onlinecite{ammeterjacs78}.}

\bibitem[{\citenamefont{Mujica et~al.}(1994)\citenamefont{Mujica, Kemp, and
  Ratner}}]{mujicajcp94II}
\bibinfo{author}{\bibfnamefont{V.}~\bibnamefont{Mujica}},
  \bibinfo{author}{\bibfnamefont{M.}~\bibnamefont{Kemp}}, \bibnamefont{and}
  \bibinfo{author}{\bibfnamefont{M.~A.} \bibnamefont{Ratner}},
  \bibinfo{journal}{J. Chem.\ Phys.} \textbf{\bibinfo{volume}{101}},
  \bibinfo{pages}{6856} (\bibinfo{year}{1994}).

\bibitem[{\citenamefont{Hansen et~al.}(2000)\citenamefont{Hansen, Nielsen,
  Brandbyge, L{\ae}gsgaarrd, Stensgaard, and Besenbacher}}]{hansenapl00}
\bibinfo{author}{\bibfnamefont{K.}~\bibnamefont{Hansen}},
  \bibinfo{author}{\bibfnamefont{S.~K.} \bibnamefont{Nielsen}},
  \bibinfo{author}{\bibfnamefont{M.}~\bibnamefont{Brandbyge}},
  \bibinfo{author}{\bibfnamefont{E.}~\bibnamefont{L{\ae}gsgaarrd}},
  \bibinfo{author}{\bibfnamefont{I.}~\bibnamefont{Stensgaard}},
  \bibnamefont{and}
  \bibinfo{author}{\bibfnamefont{F.}~\bibnamefont{Besenbacher}},
  \bibinfo{journal}{Appl.\ Phys.\ Lett.} \textbf{\bibinfo{volume}{77}},
  \bibinfo{pages}{708} (\bibinfo{year}{2000}).

\bibitem[{\citenamefont{Mehrez et~al.}(2002)\citenamefont{Mehrez, Wlasenko,
  Larade, Taylor, Gr{\"u}tter, and Guo}}]{mehrezprb02}
\bibinfo{author}{\bibfnamefont{H.}~\bibnamefont{Mehrez}},
  \bibinfo{author}{\bibfnamefont{A.}~\bibnamefont{Wlasenko}},
  \bibinfo{author}{\bibfnamefont{B.}~\bibnamefont{Larade}},
  \bibinfo{author}{\bibfnamefont{J.}~\bibnamefont{Taylor}},
  \bibinfo{author}{\bibfnamefont{P.}~\bibnamefont{Gr{\"u}tter}},
  \bibnamefont{and} \bibinfo{author}{\bibfnamefont{H.}~\bibnamefont{Guo}},
  \bibinfo{journal}{Phys.\ Rev.\ B} \textbf{\bibinfo{volume}{65}},
  \bibinfo{pages}{195419} (\bibinfo{year}{2002}).

\bibitem[{\citenamefont{Emberly and Kirczenow}(1998)}]{emberlyprl98}
\bibinfo{author}{\bibfnamefont{E.}~\bibnamefont{Emberly}} \bibnamefont{and}
  \bibinfo{author}{\bibfnamefont{G.}~\bibnamefont{Kirczenow}},
  \bibinfo{journal}{Phys.\ Rev.\ Lett.} \textbf{\bibinfo{volume}{81}},
  \bibinfo{pages}{5205} (\bibinfo{year}{1998}).

\bibitem[{\citenamefont{Landau and Lifshitz}(1958)}]{landau}
\bibinfo{author}{\bibfnamefont{L.~D.} \bibnamefont{Landau}} \bibnamefont{and}
  \bibinfo{author}{\bibfnamefont{E.~M.} \bibnamefont{Lifshitz}},
  \emph{\bibinfo{title}{Statistical Physics}}, vol.~\bibinfo{volume}{5} of
  \emph{\bibinfo{series}{Course of Theoretical Physics}}
  (\bibinfo{publisher}{Pergamon}, \bibinfo{address}{Bristol, Great Britain},
  \bibinfo{year}{1958}).

\bibitem{xiaojacs04suppl}
\bibinfo{note}{These parameters are chosen in accordance with the supplementary information to
  Ref.~\onlinecite{xiaojacs04}, freely available on the internet.}

\bibitem[{\citenamefont{Emberly and Kirczenow}(2002)}]{emberlycp02}
\bibinfo{author}{\bibfnamefont{E.~G.} \bibnamefont{Emberly}} \bibnamefont{and}
  \bibinfo{author}{\bibfnamefont{G.}~\bibnamefont{Kirczenow}},
  \bibinfo{journal}{Chem.\ Phys.} \textbf{\bibinfo{volume}{281}},
  \bibinfo{pages}{311} (\bibinfo{year}{2002}).

\bibitem[{\citenamefont{Li et~al.}(2006)\citenamefont{Li, He, Hihath, Xu,
  Lindsay, and Tao}}]{lijacs06}
\bibinfo{author}{\bibfnamefont{X.}~\bibnamefont{Li}},
  \bibinfo{author}{\bibfnamefont{J.}~\bibnamefont{He}},
  \bibinfo{author}{\bibfnamefont{J.}~\bibnamefont{Hihath}},
  \bibinfo{author}{\bibfnamefont{B.}~\bibnamefont{Xu}},
  \bibinfo{author}{\bibfnamefont{S.~M.} \bibnamefont{Lindsay}},
  \bibnamefont{and} \bibinfo{author}{\bibfnamefont{N.}~\bibnamefont{Tao}},
  \bibinfo{journal}{J. Am.\ Chem.\ Soc.} \textbf{\bibinfo{volume}{128}},
  \bibinfo{pages}{2135} (\bibinfo{year}{2006}).

\bibitem{alkanenote}
\bibinfo{note}{The method outlined in the present paper gives a conductance of
  $1.3\times 10^{-4}g_0$ for a hollow-hollow gold--octanedithiol--gold
  junction, in good agreement with the experimental results of Ref.~\onlinecite{lijacs06}.}

\bibitem{kruegerprl02}
\bibinfo{author}{\bibnamefont{D.}~\bibnamefont{Kr{\"u}ger}},
\bibinfo{author}{\bibnamefont{H.}~\bibnamefont{Fuchs}},
\bibinfo{author}{\bibnamefont{R.}~\bibnamefont{Rosseau}},
\bibinfo{author}{\bibnamefont{D.}~\bibnamefont{Marx}},
\bibinfo{author}{\bibnamefont{M.}~\bibnamefont{Parrinello}},
\bibinfo{journal}{Phys.\ Rev.\ Lett.} \textbf{\bibinfo{\volume}{89}},
\bibinfo{pages}{186402} (\bibinfo{year}{2002}).

\bibitem{ductilegold}
\bibinfo{note}{Here, we assume that small changes in the top-bound
  geometry's Au-S-C bond angle are energetically favored over forming the
  monatomic chain-forming process of 
  Ref.~\onlinecite{kruegerprl02}.
%  D. Kr{\"u}ger, H. Fuchs, R. Rousseau,
%  D. Marx, M. Parrinello \emph{Phys.\ Rev.\ Lett.\ } {\bf 89}, 186402
%  (2002). 
If not, both geometries would exhibit the unsloped conductance
  plateaus seen in experiment.}

\bibitem{stretchingnote}
\bibinfo{note}{The greater length of stretching sometimes observed in the
  experiments is presumably due to motion within the larger experimental
  system, outside the scope of the present study.}

\bibitem[{geo()}]{geometry}
\bibinfo{note}{The effect of bond geometry on molecular transmission has been
  well studied. See, for example, E. G. Emberly and G. Kirczenow, Phys.\
  Rev.\ B {\bf 58}, 10911 (1998); M. Di Ventra, S. T. Pantelides, and N. D.
  Lang, Phys.\ Rev.\ Lett.\ {\bf 84}, 979 (2000); P. E. Kornilovitch
  and A. M. Bratkovsky, Phys.\ Rev. B {\bf 64}, 195413 (2001); C. Toher,
  A. Filippetti, S. Sanvito, and K. Burke, Phys.\ Rev.\ Lett.\ {\bf
  95}, 146402 (2005).}

\bibitem[{\citenamefont{Dalgleish and
  Kirczenow}(2006{\natexlab{a}})}]{dalgleishnl06}
\bibinfo{author}{\bibfnamefont{H.}~\bibnamefont{Dalgleish}} \bibnamefont{and}
  \bibinfo{author}{\bibfnamefont{G.}~\bibnamefont{Kirczenow}},
  \bibinfo{journal}{Nano Lett.} \textbf{\bibinfo{volume}{6}},
  \bibinfo{pages}{1274} (\bibinfo{year}{2006}{\natexlab{a}}).

\bibitem[{\citenamefont{Dalgleish and
  Kirczenow}(2006{\natexlab{b}})}]{dalgleishprb06}
\bibinfo{author}{\bibfnamefont{H.}~\bibnamefont{Dalgleish}} \bibnamefont{and}
  \bibinfo{author}{\bibfnamefont{G.}~\bibnamefont{Kirczenow}},
  \bibinfo{journal}{Phys.\ Rev.\ B} \textbf{\bibinfo{volume}{73}},
  \bibinfo{pages}{245431} (\bibinfo{year}{2006}{\natexlab{b}}).

\end{thebibliography}
\end{document}